\documentclass[twoside,leqno,twocolumn]{article}

\usepackage[letterpaper]{geometry}
\usepackage{authblk}
\usepackage{ltexpprt}
\usepackage{amsmath}
\usepackage{amssymb}
\usepackage{graphicx}
\usepackage{caption}
\usepackage{subcaption}
\usepackage{float}
\usepackage{url}
\usepackage{hyperref}

\usepackage{algorithm}
\usepackage{algpseudocode}

\newcommand{\etal}{et al.\ }

\newcommand{\predI}{\hat{\cI}}

\newcommand{\predw}{\hat{\alpha}}

\newcommand{\cA}{{\cal A}}

\newcommand{\cC}{{\cal C}}

\newcommand{\cI}{{\cal I}}

\newcommand{\C}{{\cal C}}

\newcommand{\E}{\mathbb{E}}
\newcommand{\R}{\mathbb{R}}

\newcommand{\Z}{\mathbb{Z}}

\newcommand{\alg}{\mathrm{ALG}}
\newcommand{\opt}{\textsc{OPT}{}}
\newcommand{\eps}{\epsilon}

\newcommand{\val}{R}

\usepackage{xcolor}

\begin{document}

\title{\Large  Using Predicted Weights for Ad Delivery\thanks{Thomas Lavastida, and Benjamin Moseley: Supported in part by a Google Research Award, an Infor Research Award, a Carnegie Bosch Junior Faculty Chair and NSF grants CCF-1824303,  CCF-1845146, CCF-1733873 and CMMI-1938909.}}

\author[1]{Thomas Lavastida}
\author[1]{Benjamin Moseley}
\author[1]{R. Ravi\thanks{This material is based upon work supported in part by the U. S. Office of Naval
Research under award number N00014-21-1-2243 and the Air Force Office of
Scientific Research under award number FA9550-20-1-0080.}}
\author[2]{Chenyang Xu\thanks{The corresponding author. Supported in part by Science and Technology Innovation 2030 –"The Next Generation of Artificial Intelligence" Major Project No.2018AAA0100902 and China Scholarship Council No.201906320329.}}
\affil[1]{Carnegie Mellon University, USA \authorcr \{tlavasti, moseleyb, ravi\}@andrew.cmu.edu}
\affil[2]{Zhejiang University, China \authorcr xcy1995@zju.edu.cn}

\date{}

\maketitle


\fancyfoot[R]{\scriptsize{Copyright \textcopyright\ 2021 by SIAM\\
Unauthorized reproduction of this article is prohibited}}





\begin{abstract} 
We study the performance of a proportional weights algorithm for online capacitated bipartite matching modeling the delivery of impression ads.
The algorithm uses predictions on the advertiser nodes to match arriving impression nodes fractionally in proportion to the weights of its neighbors.   This paper gives a thorough empirical study of the performance of the algorithm on a data-set of ad impressions from Yahoo! and shows its superior performance  compared to natural baselines such as a greedy water-filling algorithm and the ranking algorithm.  

The proportional weights algorithm has recently received interest in the theoretical literature where it was shown to have strong guarantees beyond the worst-case model of algorithms augmented with predictions.  We extend these results to the case where the advertisers' capacities are no longer stationary over time.
 Additionally, we show the algorithm  has near optimal performance in the random-order arrival model when the number of impressions and the optimal matching are sufficiently large. 


\end{abstract}


\section{Introduction}\label{sec:intro}

There has been recent interest in augmenting online algorithms with machine-learned prediction.   This line of work has lead to new models of algorithm analysis for going beyond worst-case analysis~\cite{MLCachingLykouris,DBLP:conf/soda/LattanziLMV20,DBLP:conf/icalp/JiangP020,anand2020customizing}. 
The theoretical models considered in these works have led to the development of new algorithms which incorporate learned parameters (i.e. predictions) along with theoretical guarantees depending on the quality of the predictions. Typically, an algorithm's performance is parameterized by the error in the prediction. With a perfect prediction, an algorithm's performance should be stronger than the best worst-case algorithm. Additionally the algorithm's performance should be robust to moderate error in the predicted parameters.


An exciting question is how close the model is to practice and how to leverage it to develop improved practical algorithms.
The goal of this paper is to show that the model is closely tied to practice for the online matching problem that arises in impression ad allocation and to demonstrate the empirical efficiency of a recently proposed algorithm based on proportional weights.

\medskip
\noindent \textbf{Capacitated Online Matching:} Capacitated online matching is a fundamental problem faced in practice and is a special case of the Adwords problem \cite{DBLP:journals/jacm/MehtaSVV07}. In the problem, there is a set of known advertisers (offline) that each have a capacity (budget). Impressions arrive online that have edges to an arbitrary subset of advertisers.  An impression can be matched (fractionally) to at most one advertiser. The goal is to match as many impressions as possible under the capacity constraints. 

It is well-known that the best deterministic (greedy) algorithm is $\frac{1}{2}$-competitive.  Allowing for randomization, the ranking algorithm is known to be $(1-\frac{1}{e})$-competitive and this is the best possible competitive ratio for any online algorithm \cite{DBLP:conf/stoc/KarpVV90}.

The best worst-case algorithm is far from optimal.  Squeezing out the best performance in practice is fundamental in applications such as the Adwords problem~\cite{DBLP:conf/sigecom/DevenurH09}.  

An emerging line of work~\cite{DBLP:conf/soda/LattanziLMV20,lavastida2020learnable,AGKK20} has suggested that perhaps better matching algorithms  exist if they use learned information about known real world matching instances.  
That is, in practice there is often lots of data available on prior matching instances (e.g. the instance from yesterday).  The idea  is that an algorithm can  use information from prior instances to perform better in future problem instances. Information learned from past data is referred to as a prediction. This work has 
the potential to offer new algorithmic ideas for solving matching instances in practice. 

We consider the proportional weights algorithm, which has been suggested in this line of work. 

\medskip
\noindent \textbf{Proportional Weights for Online Matching:} This paper considers a recently proposed proportional weights algorithm for online matching.  The algorithm was first proposed by  Agrawal et al.~\cite{PropMatchAgrawal} and further developed in~\cite{DBLP:conf/soda/LattanziLMV20,lavastida2020learnable}.

Let $G = (I,A,E)$ be a bipartite graph with capacities $C \in \Z_+^A$ on $A$.  
We refer to $I$ as impressions and $A$ as advertisers.  
We use $m = |I|$ and $n = |A|$ for the number of impressions and advertisers.  
Each advertiser (impression) has a subset $N_a$ ($N_i$) of neighbors in the graph.  
In this paper we consider the fractional matching problem that is  represented by the following linear program\footnote{The more general AdWords problem has an objective coefficient for each allocated impression representing different values of an impression for different advertisers.}.

\begin{equation} \label{eqn:intro_matching_lp}
    \begin{array}{ccc}
         \max & \displaystyle \sum_{ia \in E} x_{ia}  \\
          \text{s.t.} & \displaystyle \sum_{a \in N_i} x_{ia} \leq 1 & \forall i \in I\\
          & \displaystyle \sum_{i \in N_a} x_{ia} \leq C_a  & \forall a \in A \\
          & x_{ia} \geq 0 & \forall ia \in E
    \end{array}
\end{equation}

The proportional weights algorithm assigns each advertiser $a \in A$ a weight $\alpha_a > 0$. The vector of weights $\alpha \in \R_+^A$ on the advertisers encodes a fractional assignment of impressions to advertisers in the following way. 
\begin{equation} \label{eqn:intro_prop_alloc}
    x_{ia}(\alpha) = \frac{\alpha_a}{ \sum_{a' \in N_i} \alpha_{a'}}
\end{equation}
That is, an impression is assigned to the advertisers in its neighborhood proportionally according to $\alpha$.    Notice that the allocation of each impression is independent of the others. Thus, if a set of weights is given a priori then the weights can be used to assign impressions online.

This algorithm  does not consider if an advertiser has been saturated and may assign extra impressions to an advertiser above its capacity.  These impressions are effectively not allocated.  In the online setting, we consider an improved version that uses the weights to assign the impression proportionally, but only among the neighborhood of advertisers that have remaining capacity.   This is the natural adaptation to the case where an advertiser becomes saturated. 

 Agrawal et al.~\cite{PropMatchAgrawal}  showed that for any $\epsilon > 0$, there exists a set of weights $\alpha \in \R_+^A$ such that the allocation given by \eqref{eqn:intro_prop_alloc} is a $(1-\epsilon)$-approximate solution to \eqref{eqn:intro_matching_lp}: the running time to arrive at such weights is inversely proportional to $\epsilon^2$.   This establishes that there exists a set of weights giving a high quality matching; notice that it is not obvious that such weights exist in the first place. 

The work of Agrawal et al.~\cite{PropMatchAgrawal} was interested in this proportional weights algorithm because they give a static assignment of impressions (order independent) as well as allowing for each impression to be assigned only knowing the neighborhood of the impressions, which is useful for distributed algorithms. 

Later these weights were considered in the algorithms augmented with predictions model \cite{DBLP:conf/soda/LattanziLMV20,lavastida2020learnable}.   
  Lavastida et al.~\cite{lavastida2020learnable} showed that predicting these weights can be used to go beyond the worst-case for online matching.  This prediction could come from computing the weights from prior instances of matching.  The work of Agrawal et al.~\cite{PropMatchAgrawal} imply the weights give near optimal performance if predicted perfectly.  \cite{lavastida2020learnable} showed that the weights are instance-robust.  Informally, this guarantees that if the weights give good performance on one instance, then  they have strong performance on similar instances.  Moreover, \cite{lavastida2020learnable} showed that if the matching instance is drawn from an unknown product distribution then  weights that give a near optimal solution are learnable in the PAC learning model (learnability). This suggests that weights can be learned in practice from prior matching instances and used on future instances to get strong online performance.

  This line of work begs the question, does the proportional weights algorithm perform well empirically? In particular, if weights are computed from prior instances of online matching can they be used to give strong performance on future instances of the matching problem in practice? For instance, can learning the weights on yesterdays data be used on today's online instance?  Understanding these questions has the potential to influence matching algorithms in numerous  applications.

\medskip
\noindent \textbf{Results:} This paper's goal is to demonstrate the empirical effectiveness of the proportional weights algorithm.  To do so, we consider a data set obtained from Yahoo!~\cite{YahooAdData} on the Adwords problem \cite{DBLP:journals/jacm/MehtaSVV07}.  This data set gives instances of advertisers and impressions over multiple days.  
We propose two algorithms utilizing predictions, one is the standard proportional weights algorithm while the other is an improved version.

We use the following strong benchmarks. One is the water-filling algorithm~\cite{DBLP:journals/tcs/KalyanasundaramP00}, which fractionally allocates the current impression so that it maximizes the minimum occupied proportion of its capacity among its neighbours.  
The other is the randomized ranking algorithm, which uses a single random ordering of the advertisers and assigns each impression to the available advertiser in its neighborhood with highest priority.
In the following results we consider several methods for setting the advertisers' capacities and the impression arrival orders.

\begin{itemize}
    \item Our improved proportional weights algorithm significantly outperforms the standard version. 
    \item Fix a single day.  Consider  weights that are computed from a random sample of the day's impressions and use them for the remaining impressions online.    The improved weights algorithm is consistently ahead of the baselines, often giving a near optimal matching.  This shows the impressions can be learned from a sample of a day's impressions and used on the remaining impressions effectively. This empirically demonstrates that the weights are learnable.
    \item Next we consider learning weights on prior days and using them on future days.  
    The impression distribution varies drastically from day to day.
    Despite this high variance, our improved weights algorithm has stronger performance than the baselines on every day tested, demonstrating robustness. 
\end{itemize}

To complement our results, we give two theoretical results.  First, we consider the weights in the random order arrival model. 
We show that the weights give a $(1-\eps)$-approximate online matching in the random order model for any constant $\eps >0$.  This algorithm uses the first $\sigma$ fraction of the arriving impressions to compute the weights (i.e. learn the weights) and then applies them to the remaining instance.  
In the result below, $m$ refers to the number of impressions arriving online while $n$ is the number of (offline) advertisers. 
This gives a similar result to prior work in the random order model~\cite{DBLP:conf/sigecom/DevenurH09}, but uses proportional weights instead of a primal-dual scheme.

\begin{theorem}
There exists an algorithm which is
$(1-\epsilon)$-competitive with probability $1-\delta$ for online matching in the random order model whenever $m = \Omega(\frac{n^2}{\sigma \epsilon^2}\log(\frac{n}{\delta}))$ and $\opt \geq \epsilon m$ for any  $\epsilon,\delta,\sigma \in (0,1)$.
\end{theorem}

Next, we give a theorem that demonstrates the robustness of the weights. This is an extension of the robustness result in~\cite{lavastida2020learnable}. 
Intuitively, we show that predicted weights are robust to modest changes in the input, including changes in advertiser capacity.

Consider a problem instance where advertiser $a$ has capacity $C_a$
and there is a set of impression types, where each type is defined by a subset of advertisers.  
Each impression of the same type has the same set of advertisers as neighbors.
Let $C_i$  be the number of impressions of type $i$.  

In Section~\ref{sec:robust} (Theorem~\ref{thm:IPW_robust}), we show that if a fixed set of weights can match at least $(1-\epsilon)\opt$  impressions on a given instance defined by $C$,
then the same weights have value at least $(1-\epsilon)\opt - 2\eta$ on any instance $C'$ where $\eta = \sum_{i} |C_i - C'_i|+ \sum_{a}|C'_a- C_a|$.  
Here $\eta$ measures the difference in the two instances.   This gives a theoretical explanation for the strong empirical performance of the weights even when advertiser capacities change and the impression volumes vary.  

\section{Related Work}\label{sec:related}

\medskip
\noindent \textbf{Algorithms with Predictions:}
In this paper we do a practical evaluation of online matching algorithms using predictions learned from past data. There has been significant recent interest in analyzing online algorithms in the presence of erroneous predictions~\cite{MLCachingLykouris,DBLP:conf/soda/Rohatgi20,IMMR20,ACEPS20,PurohitNIPS,DBLP:conf/icml/GollapudiP19,anand2020customizing,DBLP:conf/soda/LattanziLMV20,BhaskarOnlineLearning20,bamas2020learning,bamas2020primaldual,AGKK20}

Antoniadis et al.~\cite{AGKK20} looks at online weighted bipartite matching problems with predictions in the random order model.  In this setting each offline node can be matched at most once and the edges are weighted.  This differs from our setting where we consider the unweighted problem with capacities on the offline nodes.

\medskip
\noindent \textbf{Data Driven Algorithm Design:}
Using past data to learn the weights for our algorithm can be seen as a case of data driven algorithm design~\cite{balcan2020datadriven,DBLP:journals/siamcomp/GuptaR17,BalcanDDKSV19}.  This line of work is concerned with using past problem instances to learn an algorithm from some class of algorithms which will perform well on future instances drawn from the same population as the past instances.  In particular, the sample complexity (i.e. the number of past instances used by the learning algorithm) is of particular importance. 

\medskip
\noindent \textbf{Practical Algorithms for Online Matching:}
In addition to the deep theoretical understanding of online matching algorithms, there has been effort to develop algorithms that work well on real data sets.  Zhou et al.~\cite{ZhouRobustMatching} develop a robust online weighted matching algorithm based on primal-dual schemes for the random order model which account for changes in the underlying distribution and evaluate it on a display ad data set.  
Ma et al.~\cite{DBLP:journals/ior/MaS20} develop an algorithm for online assortment optimization and evaluate it on data from a hotel chain, but their emphasis is on revenue maximization rather than capacity allotment.
Chen et al.~\cite{ChenRealTimeBidding} develop a real-time bidding algorithm for display ad allocation and evaluate it on a proprietary display ad data set; their methods closely mirror the water-filling algorithm we study.

\medskip
\noindent \textbf{Random Order Model:}
There has been a line of work in analyzing online algorithms in the random order model for matching problems and more general packing integer linear programs~\cite{DBLP:conf/sigecom/DevenurH09,DBLP:conf/esa/FeldmanHKMS10,DBLP:journals/mor/MolinaroR14,DBLP:journals/ior/AgrawalWY14,DBLP:journals/jacm/DevanurJSW19,DBLP:journals/siamcomp/KesselheimRTV18,DBLP:journals/mor/GuptaM16}.  These algorithms are usually based off of some sort of primal-dual approach.  We give an alternative approach for online capacitated matching in the random order model based on using proportional weights.

\section{Preliminaries}

Recall the capacitated online matching problem represented by the linear program~\eqref{eqn:intro_matching_lp}.
In the online version of the problem, only the advertisers and their capacities are initially known to the algorithm.  The impressions $I$ (along with their neighbors) are revealed to the algorithm one at a time and it must commit to an assignment $\{x_{ia}\}_{a \in N_i}$ satisfying the constraints in \eqref{eqn:intro_matching_lp}.  We consider the case when the set of impressions $I$ are decided by an adversary, but revealed to the algorithm in random order.

\medskip \noindent
\textbf{Proportional Weights:} We consider fractional solutions to \eqref{eqn:intro_matching_lp} parameterized by weights $\alpha \in \R_+^A$, using the proportional allocation scheme in \eqref{eqn:intro_prop_alloc}.

Observe that given a fixed set of weights $\alpha$, the allocation $x_{ia}(\alpha)$ is order independent and thus is suitable for the online setting.  Next, note that $x_{ia}(\alpha)$ always satisfies the first constraint of \eqref{eqn:intro_matching_lp} with equality, but it may not satisfy the second constraint.  Any reasonable way of decreasing the allocation to satisfy the second constraint suffices, so we define $R_a(\alpha) = \min\{ \sum_{i\in I} x_{ia}(\alpha), C_a \}$ to be $a$'s contribution to the size of the fractional matching.  We utilize the following theorem due to Agrawal et al.~\cite{PropMatchAgrawal}.
\begin{Definition}
For $T \in \Z_+$ and $\epsilon > 0 $, define $\cA(T,\epsilon) = \{\alpha \in \R_+^A \mid \alpha_a = (1+\epsilon)^k, k \in [T] \}$.
\end{Definition}

\begin{theorem}[\cite{PropMatchAgrawal}] \label{thm:prop_weights}
For any bipartite graph $G = (I,A,E)$, capacities $C \in \Z_+^A$, and $\epsilon >0$ there exists $T \in \Z_+$ and $\alpha \in \cA(T,\epsilon)$ such that 
\[ \sum_{a \in A} R_a(\alpha) \geq (1-\epsilon) \opt. \]  
In particular, we can take $T = O(\frac{1}{\epsilon^2}\log(\frac{n}{\epsilon}))$. Moreover, there is a polynomial time algorithm which computes $\alpha$.
\end{theorem}

\section{Proportional Weights under Random Orders}
\label{sec:random_order}

This section considers the capacitated online matching problem in the random order model and shows that the proportional weights are learnable in this model. Under some mild assumptions, the weights computed by the first small portion of impressions can obtain a good performance for the whole instance.  

We first state the definition of the random order model.
Suppose that $I = \{i_1,i_2,\ldots,i_m\}$ and let $\gamma : [m] \to [m]$ be a permutation.  Denote $I(\gamma)$ to be the sequence $(i_{\gamma(1)},i_{\gamma(2)},\ldots,i_{\gamma(m)})$, i.e. consider revealing the impressions $I$ in the order induced by $\gamma$.  
If $\gamma$ is a permutation drawn uniformly at random and $\delta \in (0,1)$, then we say that an algorithm is $c$-competitive with high probability if for all impression sets $I$:
\begin{equation} \label{eqn:competitive_whp}
    \Pr[\alg(I(\gamma)) \geq c \opt(I)] \geq 1-\delta.
\end{equation}
where $\alg(I(\gamma))$ is the size of the matching when $I(\gamma)$ is given as input to the online algorithm and $\opt(I)$ is the optimal value of \eqref{eqn:intro_matching_lp}.  Note that $\opt(I)$ does not depend on the ordering $\gamma$.   We will use $\opt$ instead of $\opt(I)$ when the context is clear.

Our algorithm takes the first $\sigma m$ impressions for some $\sigma \in (0,1)$ and reduces the capacity of each advertiser by a $\sigma$ factor, then computes proportional weights $\alpha$ using the algorithm of Theorem~\ref{thm:prop_weights}.  


Denote the first $\sigma m$ impressions by $S \subseteq I$.
Let $R_a(\alpha,S) = \min \{\sum_{i \in S} x_{ia}(\alpha),\frac{|S|}{m} C_a\}$.  Similarly, let $\opt(S)$ be the size of a maximum cardinality matching on the graph $G' = (S,A,E)$ with capacities $C'_a = \frac{|S|}{m} C_a$.  We think of $\sum_a R_a(\alpha,S)$ as the value obtained by weights $\alpha$ on an instance restricted to the impressions in $S$ and the capacities scaled down appropriately.

\begin{algorithm}
\caption{Proportional Weights in the Random Order Model \label{alg:weights_random_order}}
\begin{algorithmic}
\State \textbf{Input: }$G=(I,A,E),C,\gamma,\sigma,\epsilon$.

\State Let $S$ be the first $\sigma m$ impressions in $I(\gamma)$

\State Set $x_{ia} = 0$ for each $i \in S, a \in N_i$

\State Set $T = \Theta(\frac{1}{\epsilon^2}\log(\frac{n}{\epsilon}))$

\State Compute weights $\alpha \in \cA(T,\epsilon)$ on $G' = (S,A,E)$ with capacities $C'_a = \sigma C_a$ $\forall a \in A$

\For{each remaining impression $i$}
    \State For each $a \in N_i$, let $x_{ia}(\alpha) = \frac{\alpha_a}{\sum_{a'\in N_i}\alpha_{a'}}$
\EndFor

%
\end{algorithmic}
\end{algorithm}

We show that this algorithm performs well when the number of impressions $m$ is large relative to the number of advertisers.

\begin{theorem} \label{thm:main_thm}
Algorithm~\ref{alg:weights_random_order} is $(1-\epsilon)$-competitive with probability $1-\delta$ for online matching in the random order model whenever $m = \Omega(\frac{n^2}{\sigma \epsilon^2}\log(\frac{n}{\delta}))$ and $\opt \geq \epsilon m$ for any  $\epsilon,\delta,\sigma \in (0,1)$.
\end{theorem}

Our analysis applies two probabilistic arguments.  First, we show that $\sum_a R_a(\alpha,S) \approx \sigma \sum_a R_a(\alpha)$ for the computed weights $\alpha$ with high probability.  This involves a union bound over all possible weights in $\cA(R,\epsilon)$.  Second, we show that $\opt(S) \approx \sigma \opt$ with high probability.  This involves a union bound over cuts in the bipartite graph $G$.  Formal versions of these two statements imply the theorem.
See Appendix~\ref{sec:proofs} for complete arguments.

\section{Robustness of Proportional Weights on Similar Instances}
\label{sec:robust}
	
	\begin{algorithm}[t]
	\caption{Proportional Weights (PW)}
	\label{alg:PW}
	\begin{algorithmic}
		\State \textbf{Input: }$G= (I ,A,E)$ where $I$ and $E$ arrive online, $\{ C_a \}_{a \in A}$, predicted weights $\{\predw\}$
		
		\While{an impression $i$ comes }
			\State For each $a\in N_i$, let $x_{ia} = \frac{\predw_a}{\sum_{a'\in N_i} \predw_{a'}}$.
	    \EndWhile
	\end{algorithmic}
	\end{algorithm}

A learning-augmented algorithm can be given directly if we can predict the proportional weights. See Algorithm~\ref{alg:PW} for the description. 
For this algorithm, Lavastida~\etal~\cite{lavastida2020learnable} give a theoretical result of its competitive ratio under an assumption about advertiser capacities. This section extends this result by relaxing that assumption.

Define an impression vector for an instance to be the $m$-dimensional vector $(\ldots,C_i,\ldots)$, where each component corresponds to the supply of each impression $i$.
In~\cite{lavastida2020learnable}, Lavastida~\etal prove that when the capacity of each advertiser is fixed, for any constant $\epsilon > 0$, the $(1-\epsilon)$-approximated weights $\predw$ for instance $\hat{\cI}$ has a competitive ratio at least $1-\epsilon-2\eta/\opt$ on instance $\cI$, where $\opt$ is the optimal value of instance $\cI$ and $\eta$ is $\ell_1$ norm between the impression vectors of these two instances. In this paper, we relax the condition that the capacity needs to be fixed, and obtain a similar theorem. Define an advertiser vector, similarly, to be the $n$-dimensional vector $(\ldots,C_a,\ldots)$, where each component corresponds to the capacity of each advertiser $a$.

\begin{theorem}\label{thm:robust}
	For any constant $\eps> 0$, with the $(1-\epsilon)$-approximated weights $\predw$ for instance $\hat{\cI}$, algorithm PW has a competitive ratio at least 
	\[1-\epsilon-\frac{2\eta}{\opt}\]
	on instance $\cI$, where $\opt$ is the optimal value of instance $\cI$ and $\eta$ is $\ell_1$ norm between the impression vectors of these two instances plus the $\ell_1$ norm between two advertiser vectors.
\end{theorem}

The basic idea of this proof is first showing the difference between the performances of weights $\predw$ in the two instances is at most $\eta$ and then proving the difference of these two instances' optimal values is also at most $\eta$ with the help of a vertex cut. The details are deferred to 
Appendix~\ref{app:roubust}. 

This simple algorithm is consistent, but not robust, which means that it will achieve good results if the prediction is accurate, but will perform badly if the prediction error is large. 
Several ideas are introduced in~\cite{lavastida2020learnable} to obtain robust algorithms with proportional weights. In this paper, instead of using the complicated algorithms in~\cite{lavastida2020learnable}, we propose a very simple algorithm called \textit{Improved Proportional Weights} (IPW) that can achieve a very competitive and robust performance in the experiments. The description is given in Alg.~\ref{alg:IPW}.
	\begin{algorithm}[t]
	\caption{Improved Proportional Weights (IPW)}
	\label{alg:IPW}
	\begin{algorithmic}
		\State \textbf{Input: }$G= (I, A,E)$ where $I$ and $E$ arrive online, $\{ C_a \}_{a \in A}$, predicted weights $\{\predw\}$
		
		\While{an impression $i$ comes }
			\State Let $N^*\subseteq N_i$ be the unfull advertisers in its neighbourhood (A full advertiser is one whose capacity is  equal to its current total allocation).
			
			\State For each $a\in N^*$, let $x_{ia} = \frac{\predw_a}{\sum_{a'\in N^*} \predw_{a'}}$.
	    \EndWhile
	\end{algorithmic}
	\end{algorithm}
	
Clearly, algorithm IPW always matches more (fractional) impressions than algorithm PW and obtains a maximal matching, whose competitive ratio is at least $1/2$. Thus, we have the following theorem.

\begin{theorem}\label{thm:IPW_robust}
	For any constant $\eps> 0$, with the $(1-\epsilon)$-approximated weights $\predw$ for instance $\hat{\cI}$, algorithm IPW has a competitive ratio at least 
	\[\max\left\{1-\epsilon-\frac{2\eta}{\opt},\frac{1}{2} \right\}\]
	on instance $\cI$, where $\opt$ is the optimal value of instance $\cI$ and $\eta$ is $\ell_1$ norm between the impression vectors of these two instances plus the $\ell_1$ norm between two advertiser vectors.
\end{theorem}
\section{Experimental Results}\label{sec:exp}

In this section, we 
validate the performance of PW and IPW on realistic data empirically. 
We investigate two main aspects of applying predicted weights in practice: 

\begin{itemize}
\item Learnability - we sample a small fraction of the data we assembled as training data (motivated by the results in Section~\ref{sec:random_order}) and observe that this provides enough information to set weights that are superior in performance to the benchmarks.
\item Robustness - we use  data obtained over several days and examine the performance of our weights from previous day(s) on the current day's set of impression arrivals, and show consistent improved performance over the benchmarks across time. 
\end{itemize}

First, we describe our data source and how we set up 
instances, such as the definition of impressions, the bipartite graph between impressions and advertisers, and how the capacities of the advertisers are set based on the supply of impressions.


\subsection{Experimental Setup}
We used the Yahoo! Search Marketing Advertiser Bid-Impression-Click data, provided as part of their Advertising and Market data~\cite{YahooAdData}\footnote{The data set is available from Yahoo! upon request.}. 
The data contains nearly 78 million records each containing an allocated advertising impression. Each record provides a 
day $d$, an anonymized account id (for the advertiser) $a$, a rank $r$, a set of anonymized keyphrases $P$, average bid $b$, the number of impressions $C_P$ of the keyphrase $P$ allocated to this advertiser, and clicks $k$ out of these impressions. A row $(d,a,r,P,b,C_p,k)$ indicates that on day $d$, advertiser $a$ was allocated $C_P$ copies of impression $P$ at (average monetary) cost $b$ and obtained $k$ clicks. 
Since the impressions are presumably made available in a ranked order of preference of placement, an advertiser's placement for each of these impressions is also accompanied by its rank. 

Our maximum cardinality matching instances are built from this data set, one for each day. 
The average bid and click data are ignored.
We first define the vertex set, then the edge set and finally the supply and capacity functions.

\paragraph{Vertex Sets:} The advertiser set $A$ is 
the set of advertiser account ids.
Keyphrases in the data are defined as a combination of a set of elementary (anonymized) keywords. We use this to construct the impression set $I$ as follows.
The keyphrase base set $S$ is obtained by taking the union of all elementary keyphrases in $P$. We count the  number of occurrences of each elementary keyphrase $p\in S$ and select the top 20 most frequent keyphrases (based on the supply of the impressions they are part of). 
Let $S^*$ denote the set of these most popular keyphrases and define the impression set $I$ to be the power set of $S^*$ excluding the empty set, i.e., $I:=2^{S^*} \setminus \emptyset$. Each vertex $i\in I$ can thus be viewed as an impression type. Different keyphrases that have the same intersection with $S^*$ are this `contracted' into the same impression vertex in this formulation. 

\paragraph{Edge Set:} Next we construct the edges. Define a mapping 
$f$ from the given keyphrase sets $\{P\}$ to the impression set $I$ by $f(P) = P \cap S^*$. For each row, add edge $(f(P),a)$ into the edge set. Additionally, for any subset $i\in I$ of $f(P)$, the edge $(i,a)$ is also added. 
Although these keyphrases might not be assigned to advertiser $a$ in this table, we assume that they are relevant and hence can be assigned to advertiser $a$\footnote{While some advertisers may only seek out very specific combinations of keyphrases in their bidding, we assume that most advertisers are interested in impressions broadly matching their keyphrases of interest in our formulation, and ignore the existence of the former type of advertisers.}.

\paragraph{Impression supply:}
Intuitively, on each day, the total impression sizes corresponding to different ranks should be the same. However, this is not the case in the provided data set, potentially due to sampling effects. Thus, for each day and for each keyphrase, we only retain the rows with the rank that contains the {\em maximum} total impression size and remove the rows with other ranks \footnote{The experimental results do not vary significantly if selecting other ranks in this process of defining the impression supply.}. 
For each vertex $i\in I$, its impression supply $C_i$ is the total size of the keyphrase sets that belongs to type $i$, i.e., $C_i = \sum_{f(P)=i} C_P$.

\paragraph{Advertiser Capacities:} 
The appropriate capacity function of the advertiser set for creating challenging 
instances is not readily apparent. 
Thus, we consider 
three ways to set advertiser capacities. 



\begin{enumerate}
    \item \textbf{Random Quota:} For each impression vertex $i$, split its size $C_i$ among its neighbourhood randomly. The capacity of each advertiser is set to be its final allocation obtained in this way.
    \item \textbf{Max-min Quota:} Sort all impressions lexicographically and process each impression one by one: for each impression, allocate its supply to its neighbourhood such that the minimum allocation in the neighborhood is maximized. 
    The capacity of each advertiser is set to be its final allocation at the end of this process.
    \item \textbf{Least-degree Quota:} For each impression, assign its size equally to the advertisers with the least degree\footnote{In the graph constructed from the data set, the proportion of the least-degree neighbours is roughly half its total neighborhood for each impression.} in its neighbourhood. 
\end{enumerate}

The random method above tries to de-correlate an impression's capacity and its neighborhood's eventual demands. The second and the third are also designed to deliberately avoid correlations between the neighborhood's demands. 



For the daily instance constructed in this way, there are roughly 4500 advertisers and $85$ impressions with non-zero sizes, with roughly 8000 edges between them that can be used to allocate a total supply of about $1.8$ million copies of these impressions.

\paragraph{Computational Setup and Methods.} We conducted the experiments\footnote{Code is available at \url{https://github.com/Chenyang-1995/PredictiveWeights}} on a machine running Ubuntu 18.04 with 12 i7-7800X CPUs and 48 GB memory. 
In the experiments, our algorithms (PW and IPW) are compared to the competitive greedy/water-filling algorithm (G)~\cite{DBLP:journals/tcs/KalyanasundaramP00} and ranking algorithm (R)~\cite{DBLP:conf/stoc/KarpVV90}.
The greedy `water-filling' algorithm (G) fractionally allocates the current impression so that the proportion of capacities of all its neighbors capacities that are filled are as equal as possible (Imagining these filled proportions to be water levels, the allocation fills the lowest levels until they all rise to include another in this set, and so on).
The ranking algorithm (R) uses a single random permutation of the advertisers to set a priority order among the neighborhoods of any arriving impressions and allocates the impression in this order.
All algorithms are implemented in Python 3.6.11. All results are averaged over 4 runs.

\begin{figure}[t]
    \centering
    \begin{subfigure}[t]{0.45\textwidth}
         \centering
         \captionsetup{justification=centering}
	     \includegraphics[width=0.7\linewidth]{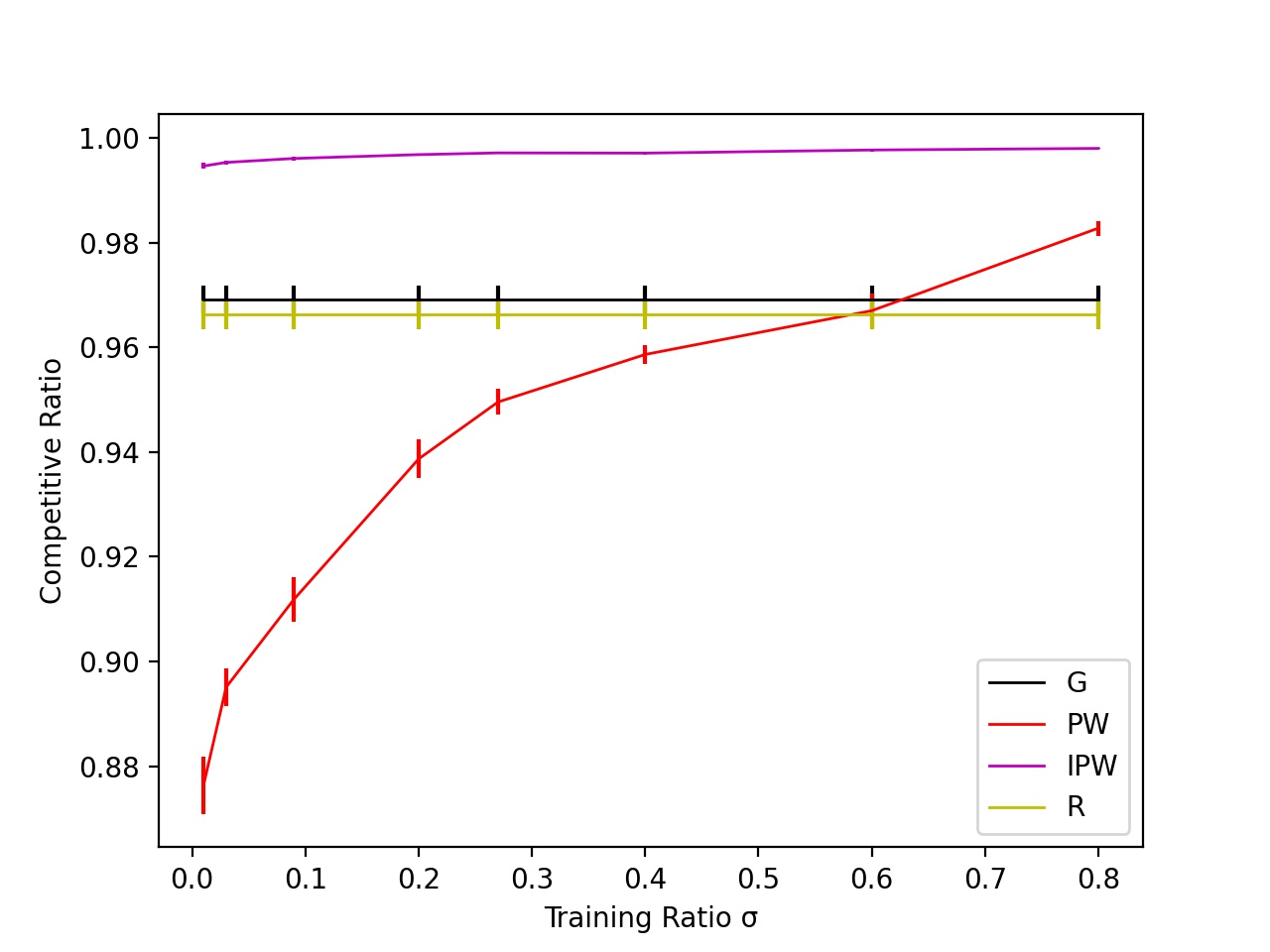}
         \caption{Random Quota}
         \label{fig:Random_Capacity_One_Day}
     \end{subfigure}
     
     \begin{subfigure}[t]{0.45\textwidth}
         \centering
         \captionsetup{justification=centering}
	     \includegraphics[width=0.7\linewidth]{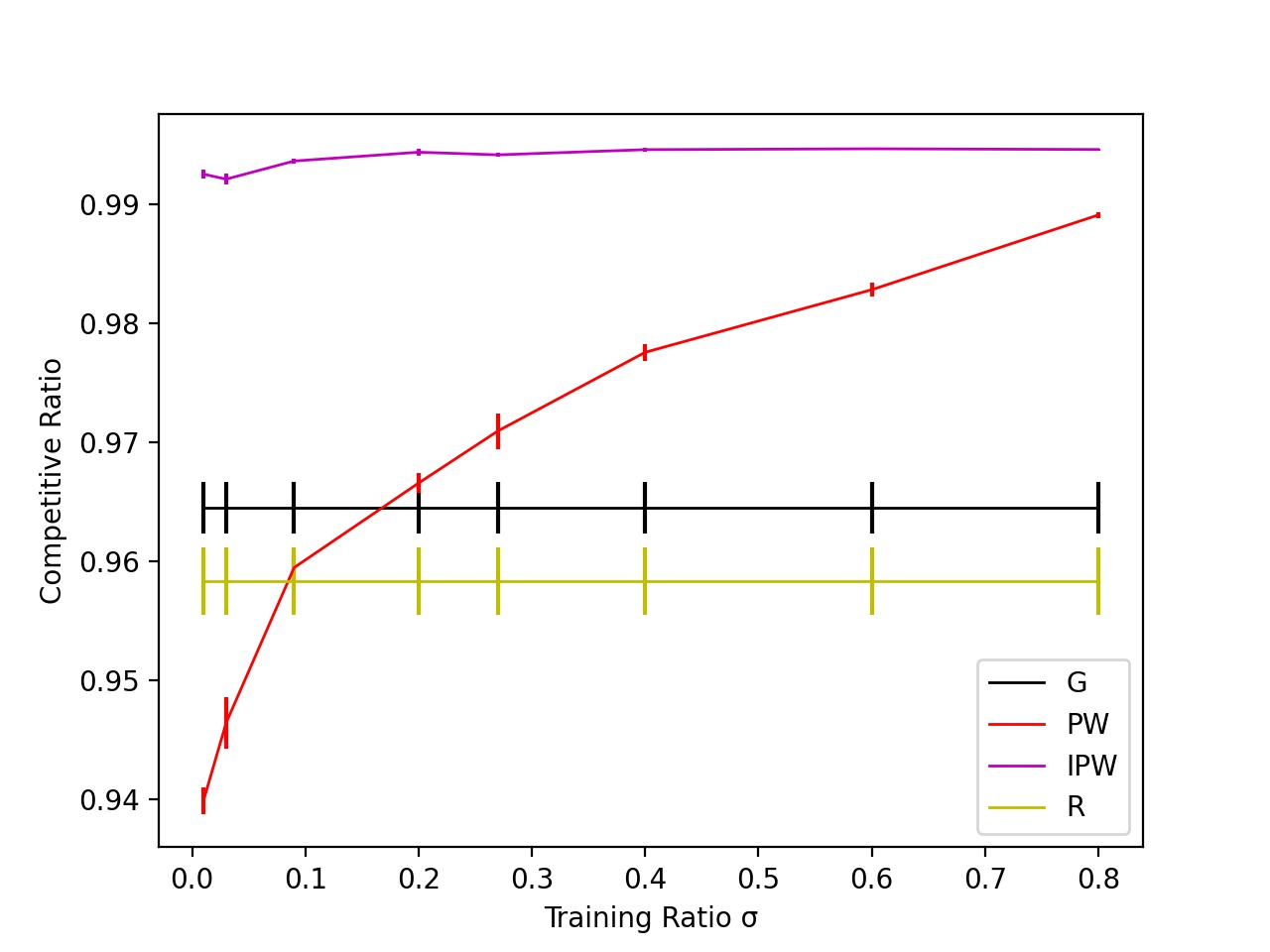}
         \caption{Max-min Quota}
         \label{fig:Min_Max_One_Day}
     \end{subfigure}
     
     \begin{subfigure}[t]{0.45\textwidth}
         \centering
         \captionsetup{justification=centering}
	     \includegraphics[width=0.7\linewidth]{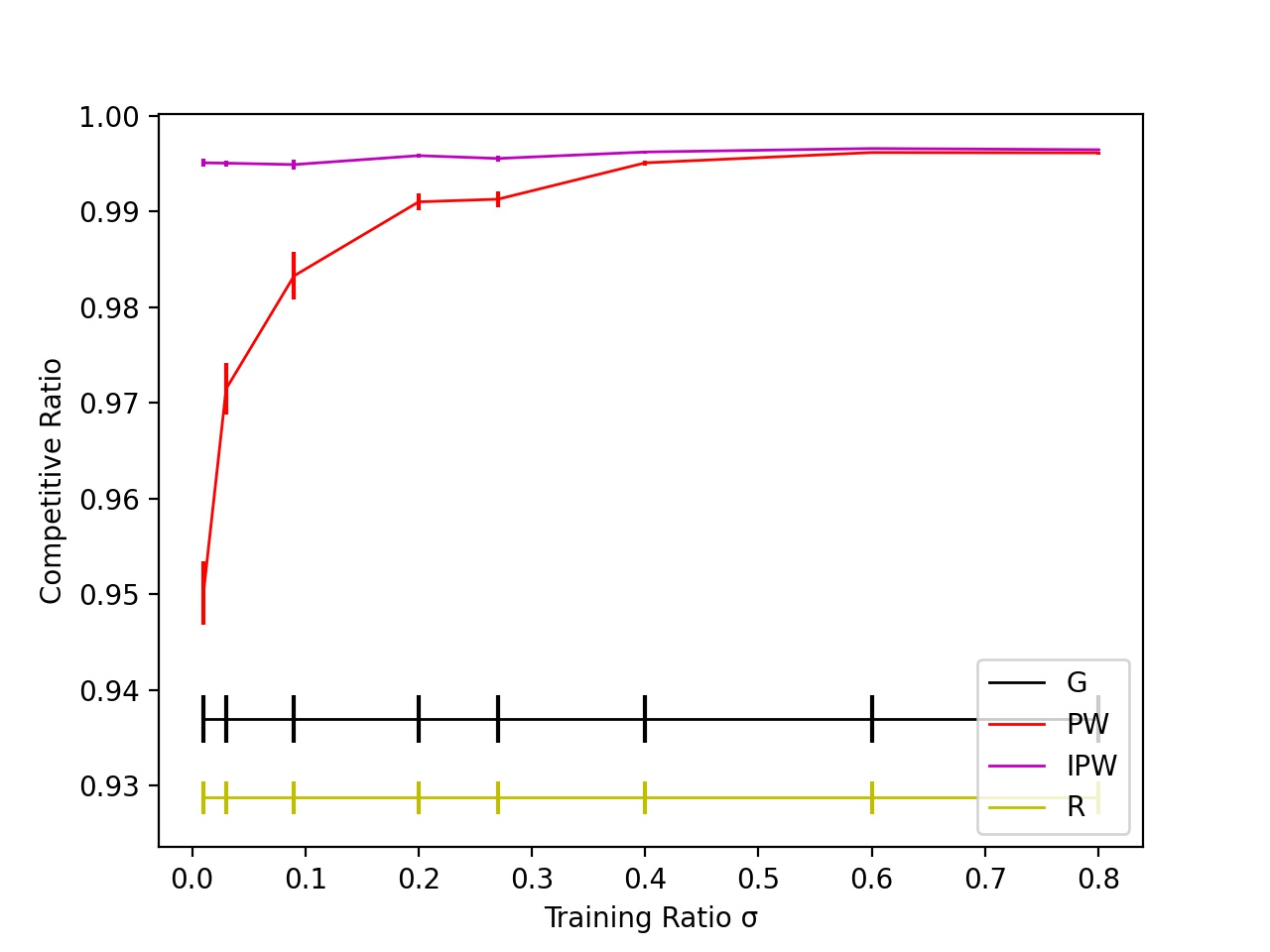}
         \caption{Least-degree Quota}
         \label{fig:Least_degree_One_Day}
     \end{subfigure}
    \caption{The performance of each algorithm on the test data when impressions arrive in a random order, plotted as a function of the training ratio.}
    \label{fig:Performance_One_Day}
\end{figure}

\begin{figure}[t]
    \centering
    \begin{subfigure}[t]{0.45\textwidth}
         \centering
         \captionsetup{justification=centering}
	     \includegraphics[width=0.7\linewidth]{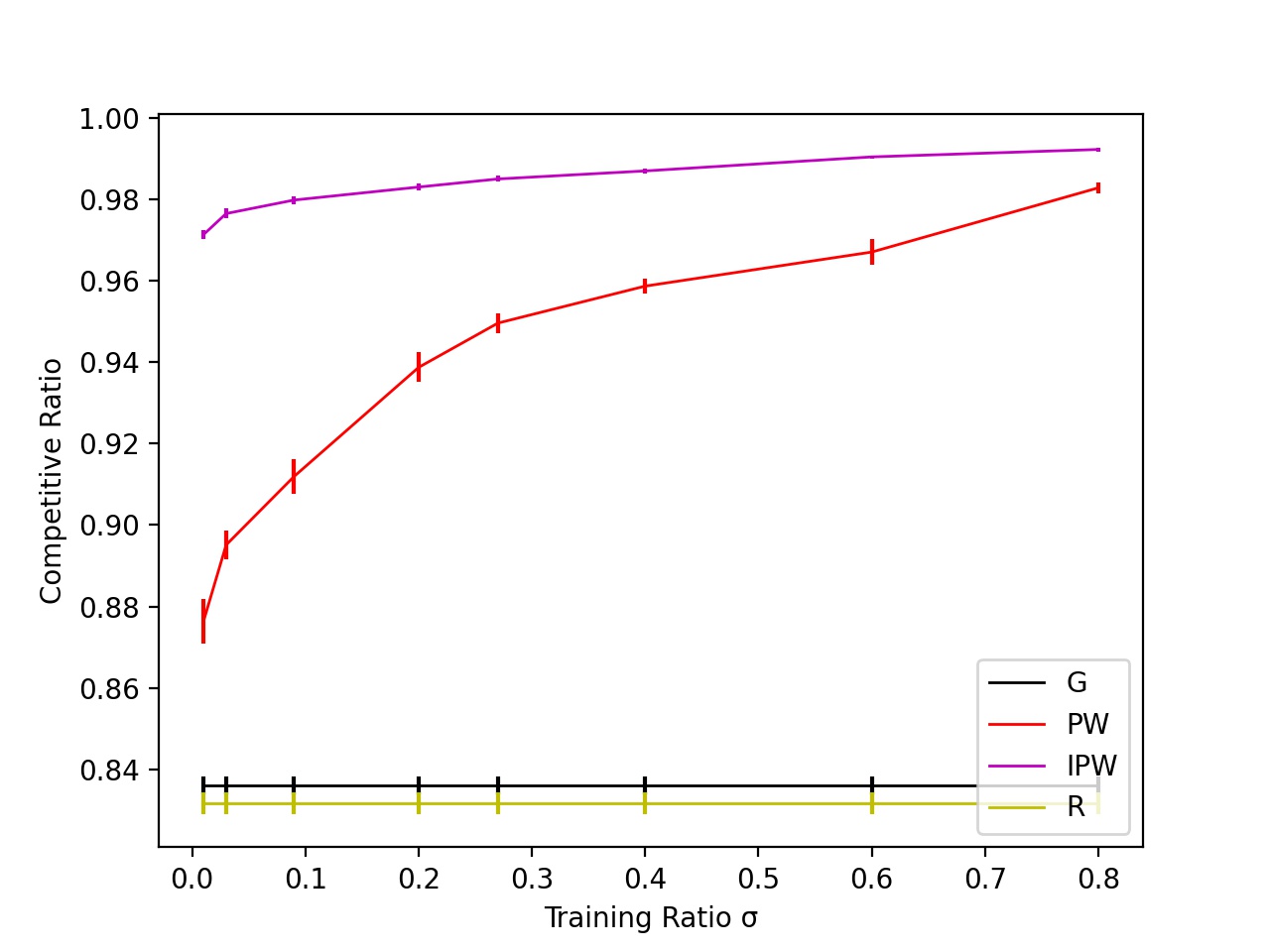}
         \caption{Random Quota}
         \label{fig:Random_Capacity_Worst_Order_One_Day}
     \end{subfigure}
     
     \begin{subfigure}[t]{0.45\textwidth}
         \centering
         \captionsetup{justification=centering}
	     \includegraphics[width=0.7\linewidth]{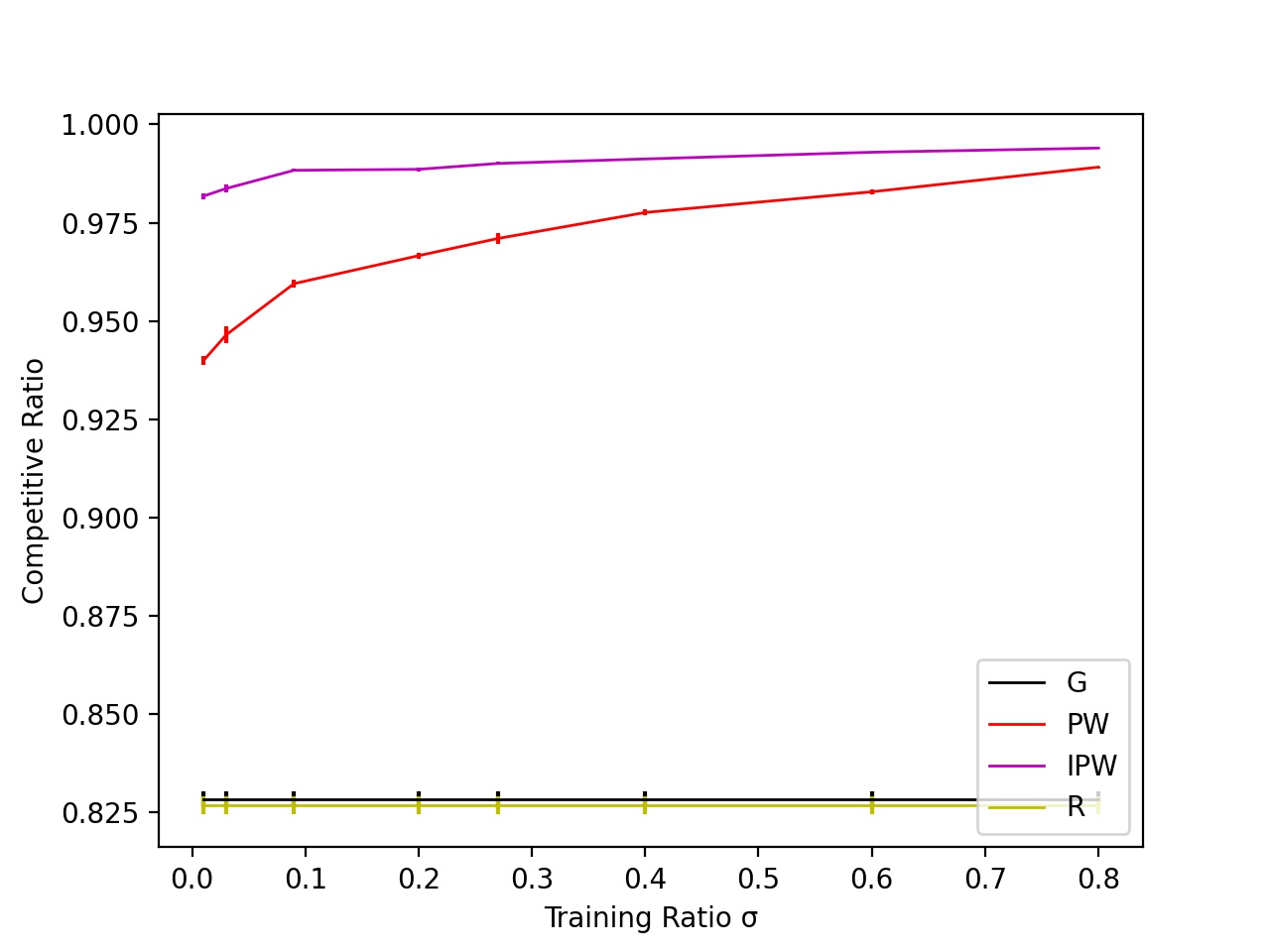}
         \caption{Max-min Quota}
         \label{fig:Min_Max_Worst_Order_One_Day}
     \end{subfigure}
     
     \begin{subfigure}[t]{0.45\textwidth}
         \centering
         \captionsetup{justification=centering}
	     \includegraphics[width=0.7\linewidth]{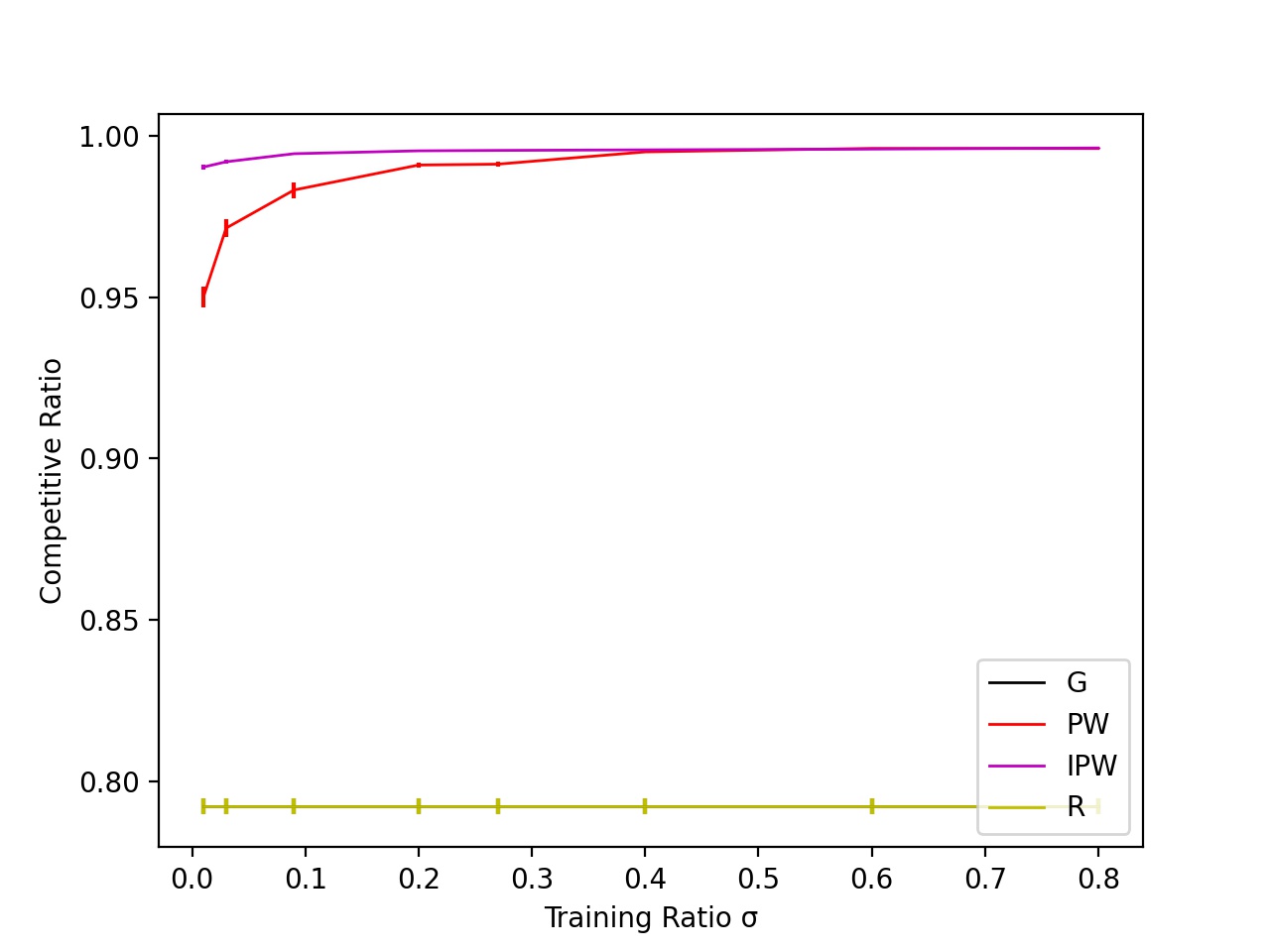}
         \caption{Least-degree Quota}
         \label{fig:Least_degree_Worst_Order_One_Day}
     \end{subfigure}
    \caption{The worst performance of each algorithm on the test data among five impression arriving orders tested, plotted as a function of the training ratio.}
    \label{fig:Performance_Worst_Order_One_Day}
\end{figure}

\begin{figure}[t]
    \centering
    \begin{subfigure}[t]{0.43\textwidth}
         \centering
         \captionsetup{justification=centering}
	     \includegraphics[width=0.7\linewidth]{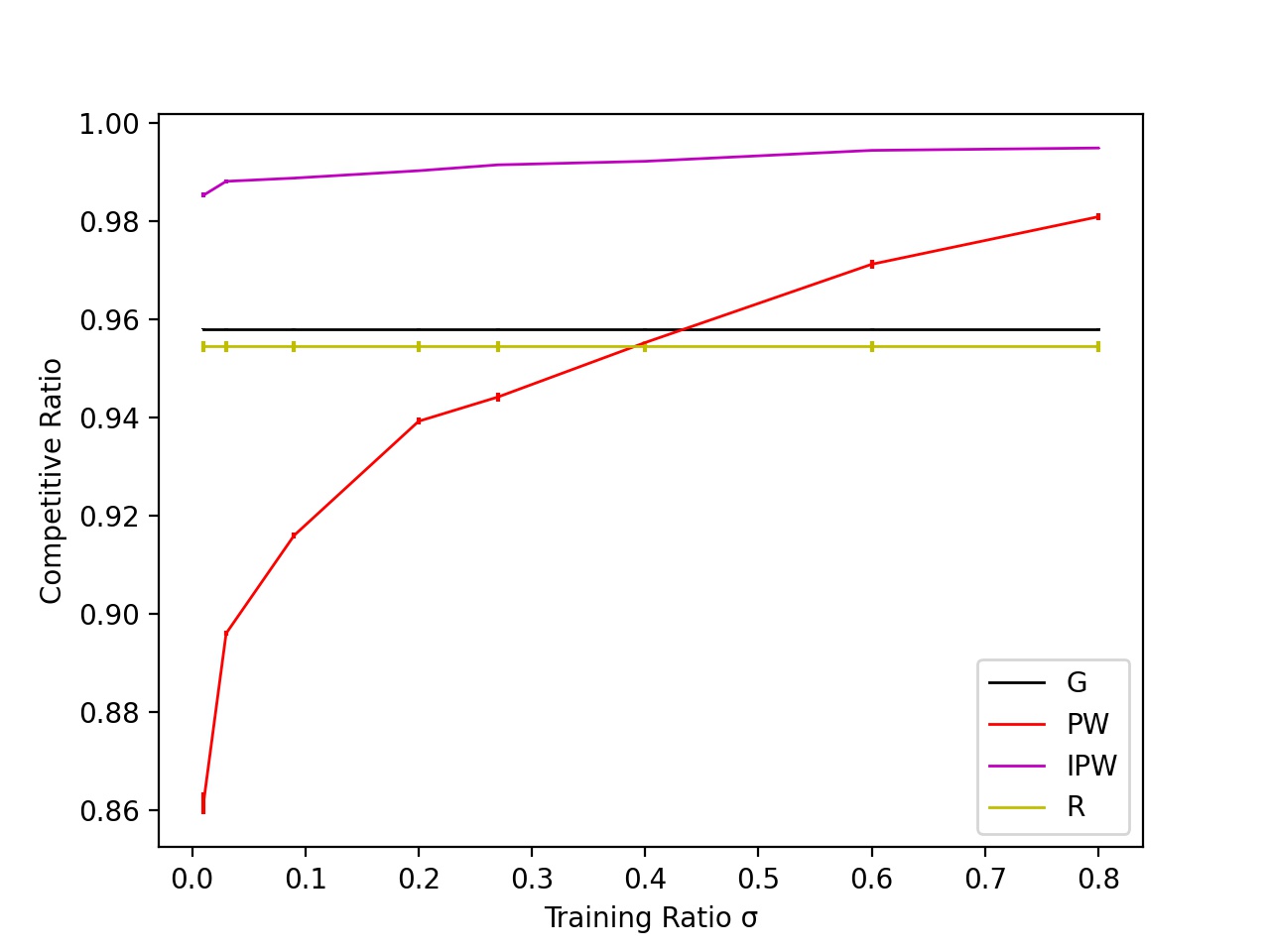}
         \caption{Random Quota}
         \label{fig:Random_Capacity_Day_Order}
     \end{subfigure}
     
     \begin{subfigure}[t]{0.43\textwidth}
         \centering
         \captionsetup{justification=centering}
	     \includegraphics[width=0.7\linewidth]{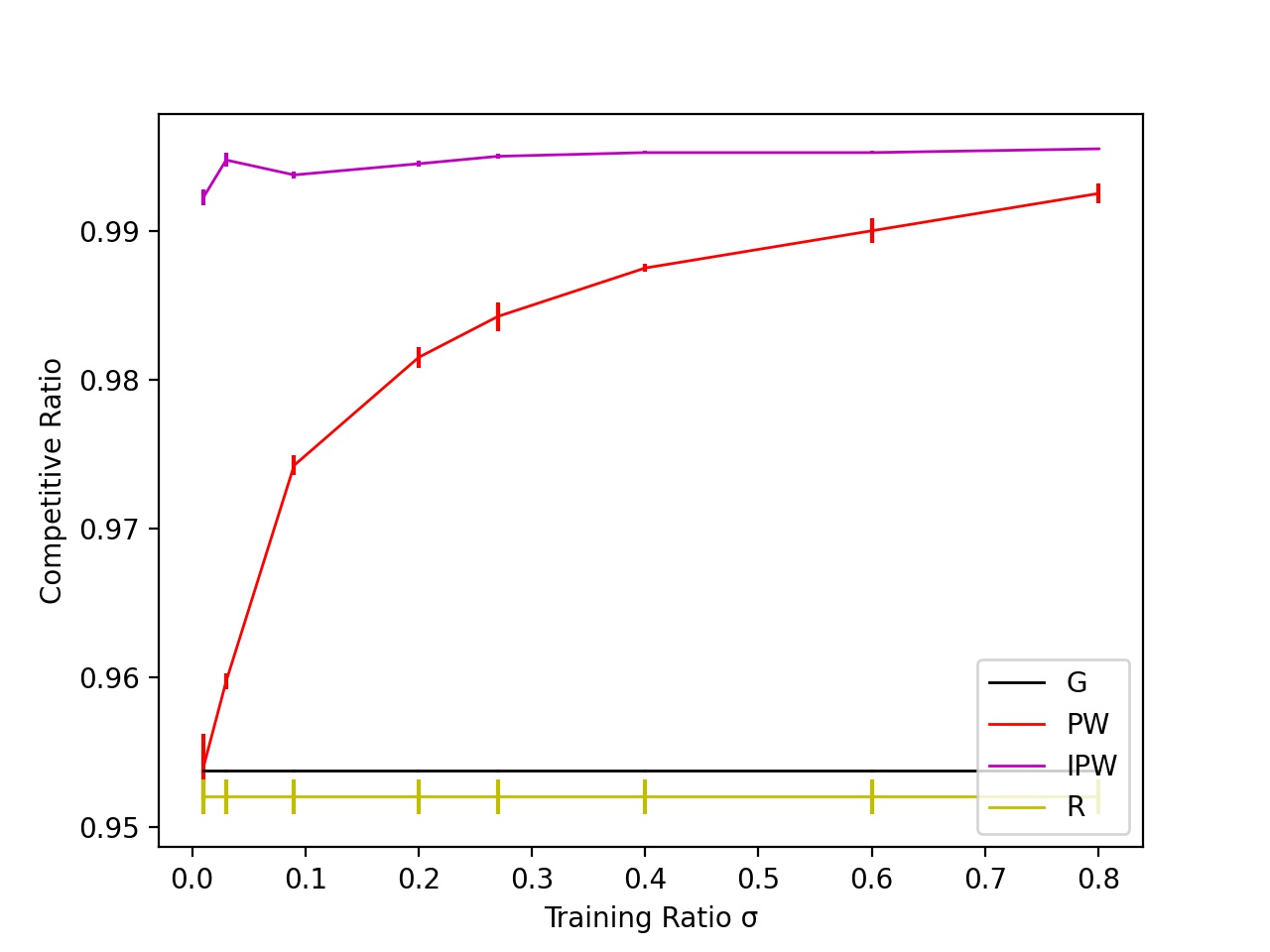}
         \caption{Max-min Quota}
         \label{fig:Min_Max_Day_Order}
     \end{subfigure}
     
     \begin{subfigure}[t]{0.43\textwidth}
         \centering
         \captionsetup{justification=centering}
	     \includegraphics[width=0.7\linewidth]{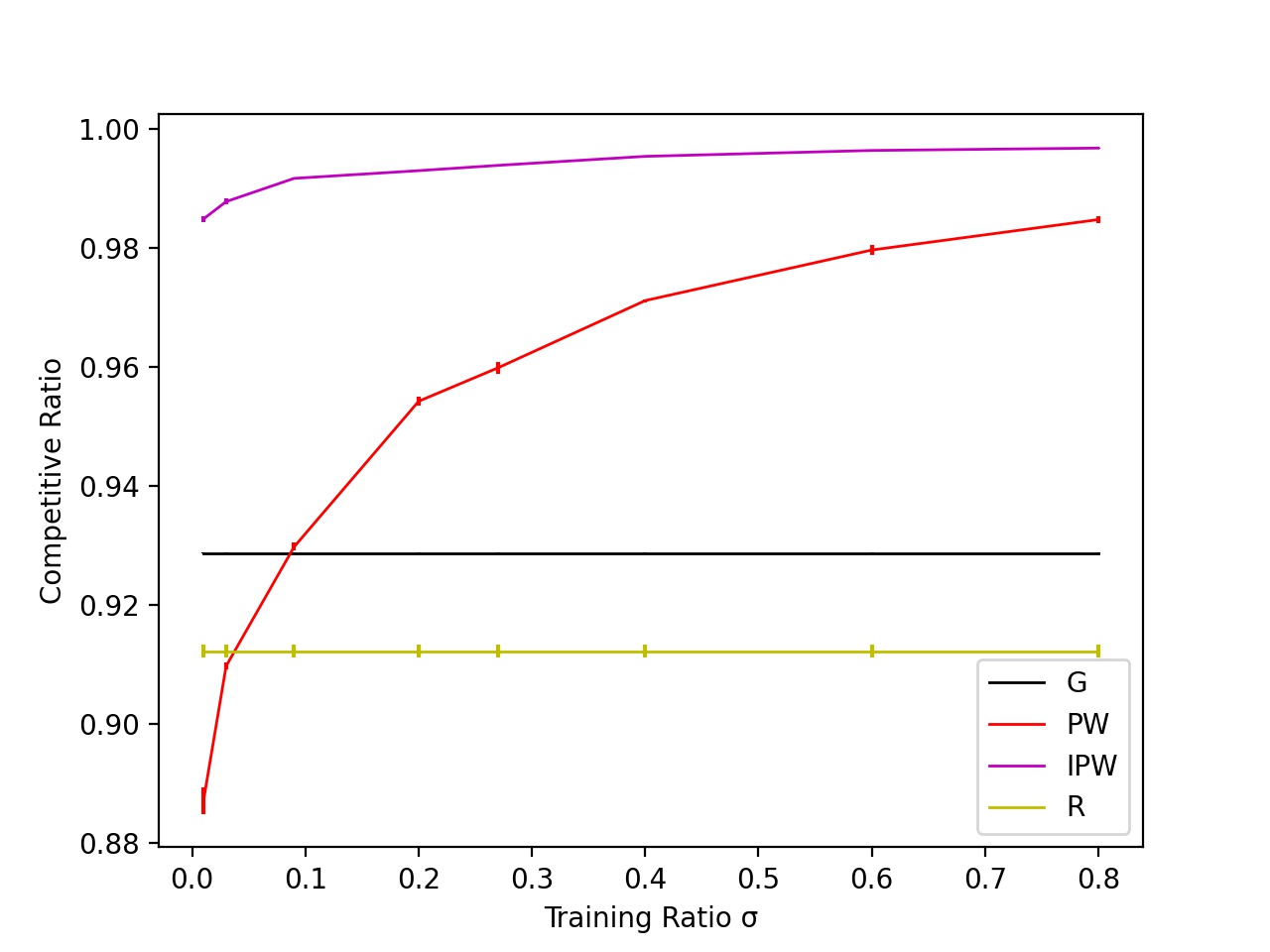}
         \caption{Least-degree Quota}
         \label{fig:Least_degree_Day_Order}
     \end{subfigure}
    \caption{The performance of each algorithm when impression arrives in a daily order, as a function of the training ratio. Impressions within a day are assumed to arrive in random order.}
    \label{fig:Performance_Day_Order}
\end{figure}

\subsection{Learnability}
To test the learnability of our algorithms, 
for each daily instance we sample a $\sigma$ proportion of impressions (for a sampling parameter $\sigma \in [0,1]$) to construct the training instance. The graph connectivity is the same as in the original data set while the advertiser capacities are constructed using these impressions and one of the three rules above.
We compute proportional weights on this training instance and use them in Algorithms~\ref{alg:PW} and~\ref{alg:IPW} for that day's whole instance. 

We investigate the performance when impressions arrive in a random order or in an adversarial order.
We measure performance by the traditional measure of competitive ratio. 
Since the instances were engineered so that all impressions are allocable, this is simply the fraction of all impressions that were assigned by each of the methods. 
Since it is hard to find the most adversarial order for each algorithm for the given impressions, we set up five different arrival orders and use the worst performance of an algorithm among these orders to approximate its performance in an adversarial order. Finally, we illustrate the performance of the algorithms across the daily order, 
which we think is closest to the order which occurs in practice.
The description of the daily order is given later.

\paragraph{Random Order.} The performance of each algorithm in random order is shown in Figure~\ref{fig:Performance_One_Day}, where each plot corresponds to one of the three quota allocation rules\footnote{Note the varying scales in the Y-axes in many of the figures.}. 
All results are obtained by taking the average performance of ten days' instances, while the performance of each day is evaluated on 4 separate runs.
The x-axis of each plot is the training ratio $\sigma$, indicating that $\sigma$ proportion of impressions are sampled to serve as the training data. 
Note that in these figures, the learned weights from the $\sigma$ proportion are evaluated on the whole data set for the same day.
As $\sigma$ increases, the baseline greedy (G) and ranking (R) algorithms' performance remain unchanged since they do not use the training data, while algorithms PW and IPW show improving performances. In Figure~\ref{fig:Random_Capacity_One_Day} and Figure~\ref{fig:Min_Max_One_Day}, algorithm PW has a worse performance than algorithm G and algorithm R initially, but obtains a better performance when $\sigma$ becomes 0.6 and 0.1 respectively. Compared to other algorithms, IPW gives the best performance. We see that even with one percent of the data ($\sigma = 0.01$), IPW outperforms the traditional online algorithms.

\paragraph{Approximating Adversarial Orders.} To study the performance of these algorithms when impressions arrive in an adversarial order, we set up following five arriving orders and check the worst performance.
\begin{enumerate}
    \item \textbf{Random Order:} Impressions arrive randomly.
    \item \textbf{$C_i$-descending Order:} Sort all impressions in the descending order of their capacities and let impressions arrive in this order.
    \item \textbf{$C_i$-ascending Order:} All impressions arrive in the opposite order of the $C_i$-descending order (i.e. non-descending order of their capacities).
    \item \textbf{$C_a$-descending Order:} Define the neighbourhood capacity of an impression to be the sum of the supplies of its neighbouring advertisers. Sort all impressions in the descending order of their neighbourhood capacities and let impressions arrive in this order.
    \item \textbf{$C_a$-ascending Order:} All impressions arrive in the opposite order of the $C_a$-descending order.
\end{enumerate}

The worst performance of each algorithm (over these five orders) is shown in Figure~\ref{fig:Performance_Worst_Order_One_Day}. 
We observe that the algorithms based on proportional weights are much more stable than other algorithms over different arrival orders (More experimental results are provided in Appendix~\ref{sec:exp_app}).
The performance of algorithms PW and IPW mostly remain unchanged because their allocation policies are unrelated to the impression arrival order, unlike greedy and ranking, and IPW remains the best. 
The performances of algorithm G and R vary significantly when the arriving order changes. Their worst performances are usually realized in the $C_i$-descending order and $C_a$-descending order respectively. 

\paragraph{Daily Order.} Finally, we show each algorithm's performance when impressions arrive in a daily order in Figure~\ref{fig:Performance_Day_Order}. 
We combine the instances on 7 days and get a stacked instance, where the vertex set is the union of vertex sets in these graphs and each vertex's capacity is the sum of its capacities in the instances.
Say that impressions arrive in daily order if all impressions on day $d$ arrive before the impressions on day $d+1$ while inside any day $d$, impressions arrive in a random order.
The results show a similar trend as in Figure~\ref{fig:Performance_One_Day} with one difference being that the performances of algorithm G and R decrease slightly. Thus, the value $\sigma$ where algorithm PW surpasses algorithm G and R becomes slightly smaller.  

\begin{figure}
    \centering
    \captionsetup{justification=centering}
    \includegraphics[width=0.9\linewidth]{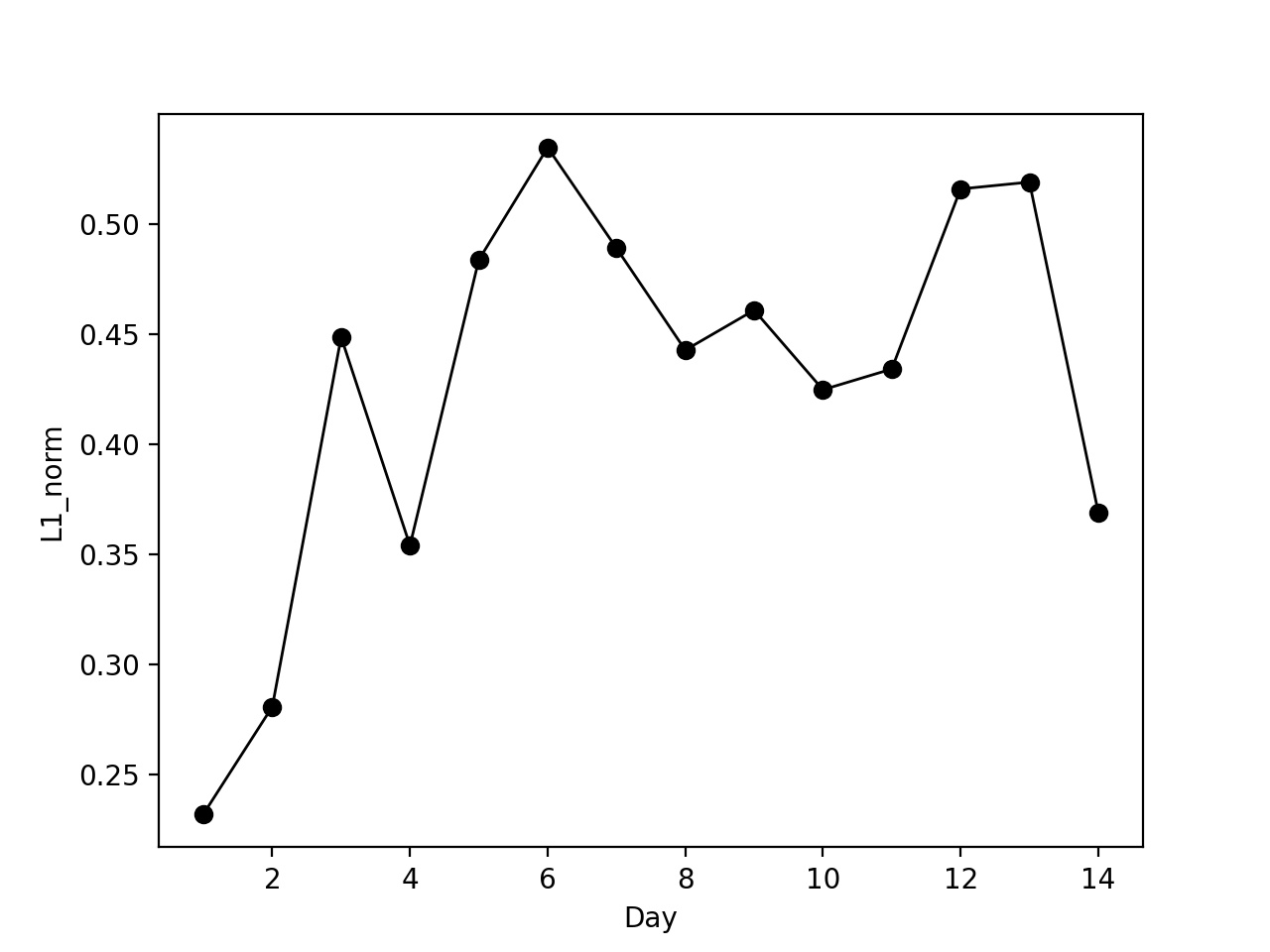}
    \caption{The $\ell_1$ norm between the impression vectors on day 0 and day $i>0$.}
    \label{fig:l1_norm}
\end{figure}

\begin{figure}[t]
    \centering
    \begin{subfigure}[t]{0.56\textwidth}
         \centering
         \captionsetup{justification=centering}
	     \includegraphics[width=0.7\linewidth]{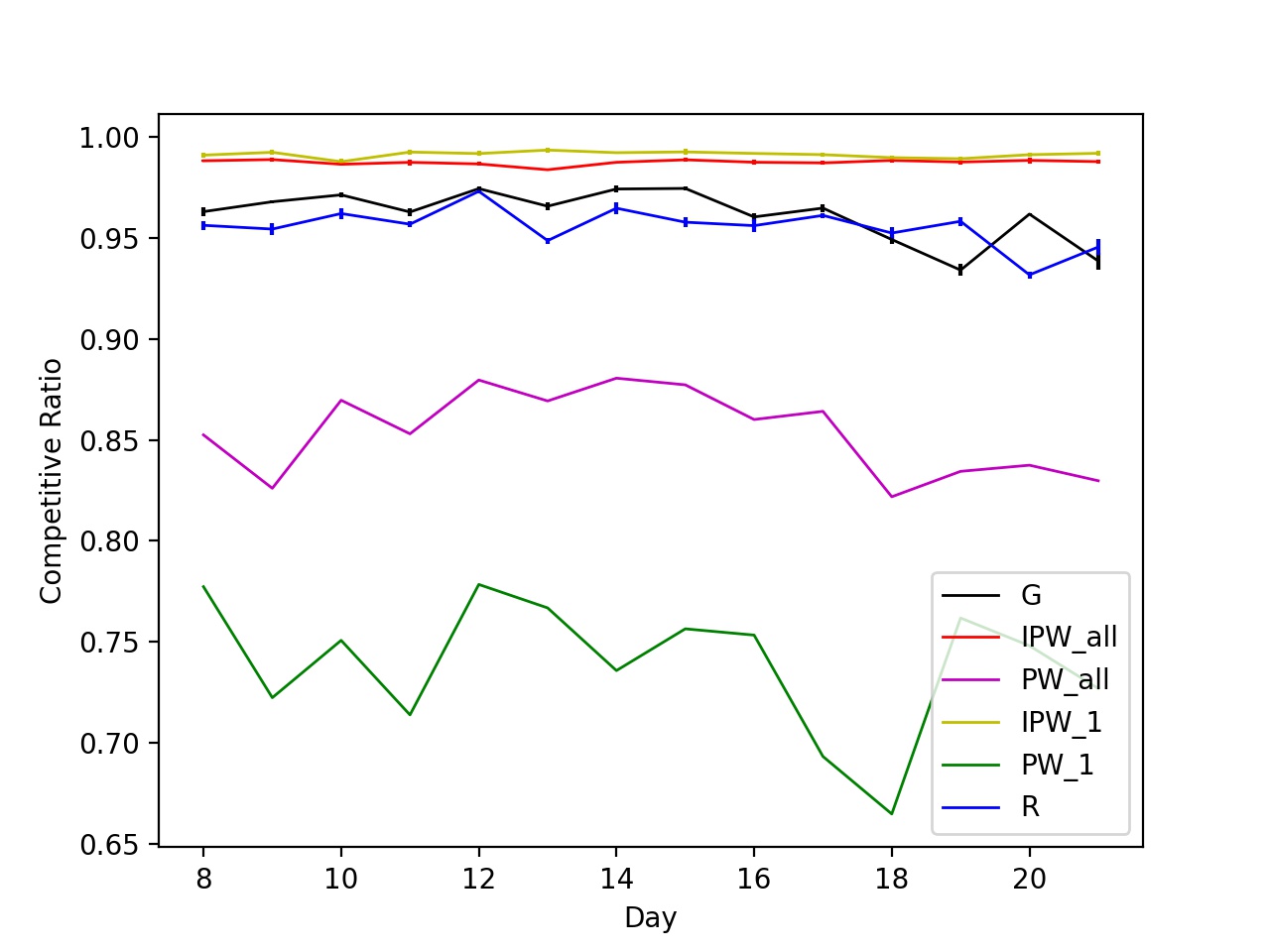}
         \caption{Max-min Quota}
         \label{fig:Min_Max_Different_Day}
     \end{subfigure}
     
     \begin{subfigure}[t]{0.56\textwidth}
         \centering
         \captionsetup{justification=centering}
	     \includegraphics[width=0.7\linewidth]{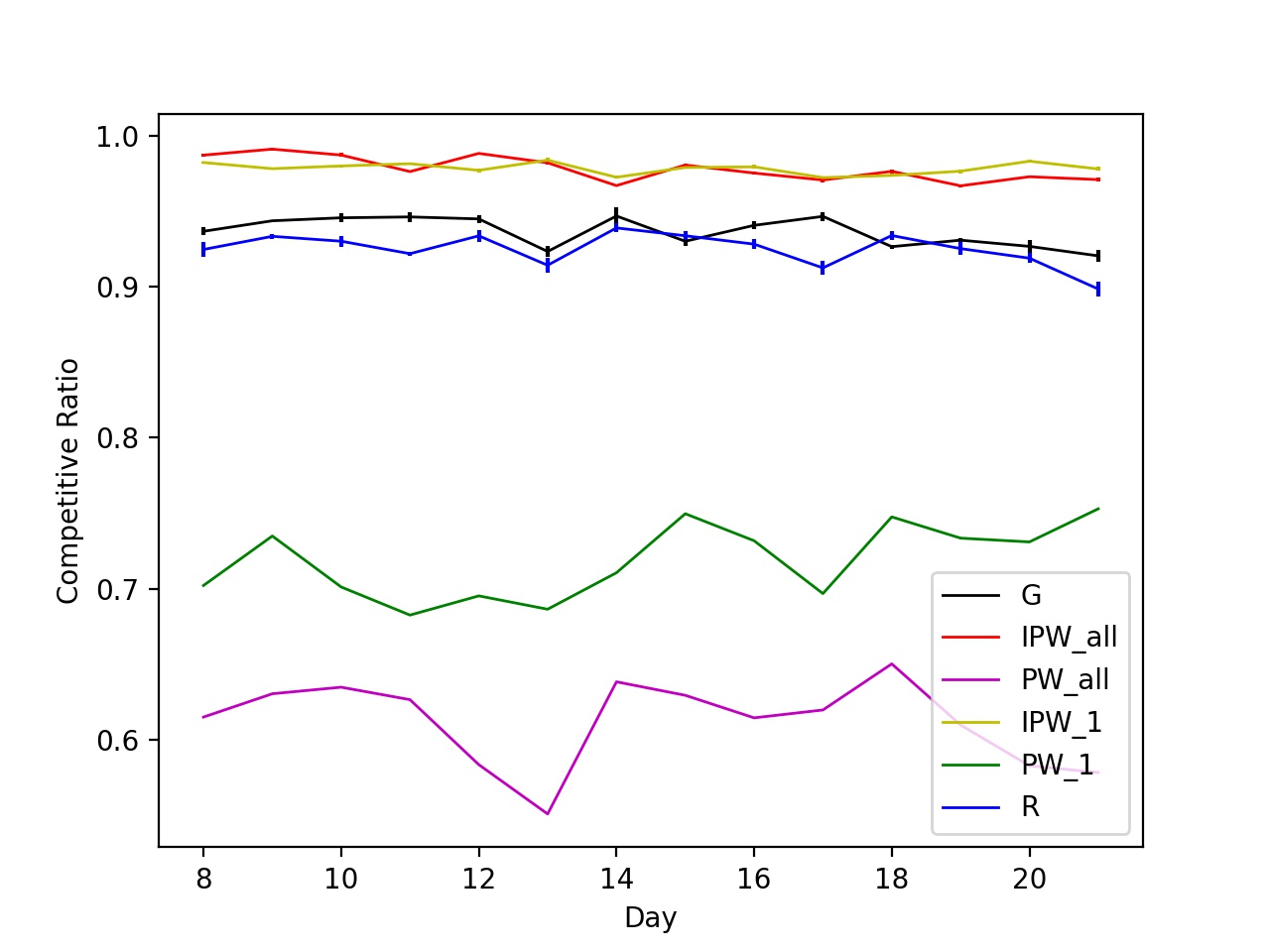}
         \caption{Least-degree Quota}
         \label{fig:Least_Degree_Different_Day}
     \end{subfigure}
    \caption{The performance of algorithms on different days where PW and IPW use weights from previous days on the current day, and the impression arrival order is random within each day. Note that the starting day is set to be day 8 in order to collect more training data for algorithm PW\_all and IPW\_all.}
    \label{fig:Performance_Different_Day}
\end{figure}

\subsection{Robustness.}

This subsection considers the robustness. As mentioned above, the instances on different days are quite different from each other. We visualize this by showing the $\ell_1$ norm between the impressions on the first day (Day 0) and the normalized impression vectors on following days, using the impressions from the given data set across the first 15 days.
In our setting, the vector has dimension $2^{20}-1$. 
As shown in Figure.~\ref{fig:l1_norm}, the difference between two days could be very large. This suggests it could be hard for  proportional weights to work well when learned on one day and used on another. Surprisingly, our experiments shows  empirically that  weights are robust across days despite these large differences in the problem instances.
Since we wish to model some level of correlation between the instances on different days, the random quota is not tested in this experiment.

The robustness experiment simulates how the proportional weights may be used in practice.  We predict the weights based on previous day(s) and use the weights to allocate impressions for a different day.   We consider either computing weights on an instance that is just the previous day or computing the weights from \emph{all} prior days impressions.  We use ``\_all'' and ``\_1'' to denote the weights of all previous days and yesterday respectively. The results are shown in Figure~\ref{fig:Performance_Different_Day} 
\footnote{If one considers the worst performance among the five orders, the performance of algorithm G and R will decrease significantly as in previous experiments, while PW and IPW remain stable.}.

As mentioned in the beginning of this section, algorithm PW will not be robust if the predicted weights are inaccurate. The performance of algorithm PW\_all  and PW\_1 imply a large error in the prediction. However, even with large prediction error IPW achieves the best performance.

\subsection{Conclusions}
We see the following trends from the experiments.

\begin{itemize}
    \item The Improved Proportional Weights algorithm is consistently the best algorithm considered, giving near optimal performance on
    all instances tested.
    \item The weights are learnable given a random sample of a problem instance.  In particular, such weights leads to the improved proportional weights algorithm having near-optimal performance.  
    \item The weights are robust to large changes in the problem instance.  
    Across days, the advertiser capacities and the supply of the impressions change by large margins. Still, the Improved Proportional Weights algorithm has strong performance.
\end{itemize}

These experiments show that (1) the proportional weights algorithm is a strong algorithm for online matching and (2) the theoretical results on the weights can be seen empirically.  This empirically shows a connection between the algorithms augmented with predictions model and practice.
We hope these results stimulate further experimental investigation of algorithms augmented with predictions.
\allowdisplaybreaks

\bibliographystyle{plain}
\bibliography{ref.bib}
\appendix


\section{Concentration Inequalities}

\begin{theorem}[Corollary 2.4 in \cite{DBLP:journals/mor/GuptaM16}]\label{thm:Bernstein}
    Let $Y=\{Y_1,...,Y_n\}$ be a set of real numbers in the interval $[0,1]$. Let $S$ be a random subset of $Y$ of size $s$ and let $Y_S=\sum_{i\in S}Y_i$. Setting $\mu=\frac{1}{n}\sum_i Y_i,$ we have that for every $\tau>0$,
    \[\Pr[|Y_S-s\mu|\geq \tau] \leq 2\exp\left(-\min \left\{ \frac{\tau^2}{8s\mu},\frac{\tau}{2} \right\} \right). \] 
\end{theorem}

\section{Proofs for the Random Order Model} \label{sec:proofs}

\begin{lemma} \label{lem:main_lemma}
Suppose that the weights $\alpha$ computed in Algorithm~\ref{alg:weights_random_order} satisfy 
\begin{enumerate}
    \item $\Pr\left[\sum_a R_a(\alpha,S) > \sigma \sum_a R_a(\alpha) + \epsilon^2 \sigma m\right] \leq \frac{\delta}{2}$
    \item $\Pr\left[\opt(S) < \sigma \opt - \epsilon^2 \sigma m \right] \leq \frac{\delta}{2}$
\end{enumerate}
Then Algorithm~\ref{alg:weights_random_order} is $(1-O(\epsilon))$-competitive with probability $1-\delta$ whenever $\opt \geq \epsilon m$.
\end{lemma}
\begin{proof}
The key fact we use is that for any $S$, by Theorem~\ref{thm:prop_weights}, $\sum_a R_a(\alpha,S) \geq (1-\epsilon) \opt(S)$ since we computed $\alpha$ using $S$.  If neither event above occurs, then we have
\[
\begin{split}
   \sum_a R_a(\alpha) &\geq \frac{1}{\sigma} \sum_a R_a(\alpha,S) - \epsilon^2 m \\
   & \geq \frac{(1-\epsilon)}{\sigma} \opt(S) - \epsilon^2m \\
    & \geq (1-\epsilon)\opt - 2\epsilon^2 m  \\
    & \geq  (1-O(\epsilon))\opt  . 
\end{split}
\]
Indeed, neither event occurs with probability at least $1-\delta$ by a union bound.
\end{proof}

Our goal is to now show that the two properties above hold whenever $m = \Omega(\frac{n^2}{\sigma \epsilon^2}\log(\frac{n}{\delta}))$.  Let's start with the first property.  In order to show this property we need the following lemma.

\begin{lemma} \label{lem:swap_min_for_sum}
Let $S$ be a subset $I$ of size $\sigma m$ and $\alpha \in \cA(T,\epsilon)$.  For any $a \in A$, if $R_a(\alpha,S) \geq \sigma R_a(\alpha)$ then $\sum_{i\in S} x_{ia}(\alpha) \geq \sigma \sum_{i \in I} x_{ia}(\alpha)$.
\end{lemma}
\begin{proof}
Fix $a \in A$ and suppose that $R_a(\alpha,S) \geq  \sigma R_a(\alpha)$. Using the definition of $R_a(\alpha,S)$ and $R_a(\alpha)$ we have
\[ 
\begin{split}
    \min \left\{\sum_{i \in S} x_{ia}(\alpha), \sigma C_a \right\} & = R_a(\alpha,S) 
     \geq \sigma R_a(\alpha) \\ & = \sigma \min \left\{ \sum_{i\in I} x_{ia}(\alpha), C_a  \right\}\\
     & =\min \left\{ \sigma \sum_{i\in I} x_{ia}(\alpha), \sigma C_a  \right\}
\end{split}
\]
A case analysis of this shows that $\sum_{i\in S} x_{ia}(\alpha) \geq  \sigma \sum_{i \in I} x_{ia}(\alpha)$.
\end{proof}

This lemma allows us to bound the probability that $R_a(\alpha,S) > (1+\epsilon) \sigma R_a(\alpha)$ by instead bounding the probability that $\sum_{i\in S} x_{ia}(\alpha) \geq (1+\epsilon)\sigma \sum_{i \in I} x_{ia}(\alpha)$, which is accomplished via standard concentration inequalities.

\begin{lemma} \label{lem:property1}
For each $\alpha \in \cA(T,\epsilon)$ and $a \in A$, if $m = \Omega(\frac{n}{\epsilon^2\sigma} \log (\frac{2n|\cA(T,\epsilon)|}{\delta}))$ then 
\[\Pr[R_a(\alpha,S) >  \sigma R_a(\alpha) + \epsilon^2\sigma  \frac{ m}{n}] \leq \frac{\delta}{2n|\cA(T,\epsilon)|}.\]
\end{lemma}
\begin{proof}
By Lemma~\ref{lem:swap_min_for_sum}, we have 
\[ \begin{split}
    \Pr& [R_a(\alpha,S) >  \sigma R_a(\alpha) + \epsilon^2\sigma  \frac{ m}{n}] \\
    & \leq \Pr[\sum_{i\in S} x_{ia}(\alpha) >  \sigma \sum_{i \in I} x_{ia}(\alpha) +\epsilon^2 \sigma \frac{m}{n}].
\end{split}\]  
Since $S$ is a random subset of $I$ of size $\sigma m$, $x_{ia} \in [0,1]$, and $\E[ \sum_{i\in S} x_{ia}(\alpha)] = \sigma \sum_{i \in I} x_{ia}(\alpha)$, we can apply Theorem~\ref{thm:Bernstein} to the right hand side to get
\[ \begin{split}
\Pr & \left[\sum_{i\in S} x_{ia}(\alpha) >  \sigma \sum_{i \in I} x_{ia}(\alpha) +\epsilon^2 \sigma \frac{m}{n} \right] \\
& \leq \exp\left( - \epsilon^2 \sigma \frac{m }{2n} \right) \leq \frac{\delta}{2n|\cA(T,\epsilon)|}
\end{split}\]
where in the last step we use the condition on $m$.
\end{proof}

We use a similar strategy for showing the second condition which regards $\opt$ and $\opt(S)$: find a quantity which can be captured as a sum and apply concentration.  In this case we use the fact that there exists a cut/vertex cover which equals the size of the maximum cardinality (fractional) matching. 
The following theorem is folklore (see e.g. Claim 1 in \cite{PropMatchAgrawal}).  For $B \subseteq A$ let $N(B) = \bigcup_{a \in B} N_a$.

\begin{theorem} \label{thm:cut}
Let $G = (I,A,E)$ be a bipartite graph with capacities $C \in Z_+^A$ on $A$.  If $\opt$ is the optimum value of \eqref{eqn:intro_matching_lp}, then for all partitions of $A$ into $A' \cup A''$ we have $\opt \leq |N(A')| + \sum_{a \in A''} C_a$.  Moreover, there exists a partition of $A$  into $A_0 \cup A_1$ such that $\opt = |N(A_1)| + \sum_{a \in A_1}C_a $.
\end{theorem}

We can apply this theorem directly to $G$ as well as to $G' = (S,A,E)$ as in Algorithm~\ref{alg:weights_random_order}.  The capacities for $G'$ are $C'_a = \sigma C_a$, so for any $S \subseteq I$ of size $\sigma m$, we get that there is a partition $A = A_0 \cup A_1$ such that $\opt(S) = |N(A_0) \cap S| + \sum_{a \in A_1} \sigma C_a$.  Now we can write $|N(A_0)\cap S| = \sum_{i \in S} Y_i$, where $Y_i = \mathbf{1}_{\{i \in N(A_0)\}}$, which will allow us to apply Theorem~\ref{thm:Bernstein}.

\begin{lemma} \label{lem:property2}
If $m = \Omega(\frac{1}{\epsilon^2\sigma}(n+ \log\frac{2}{\delta}))$, then \[ \Pr[\opt(S) < \sigma \opt - \epsilon^2 \sigma m] \leq \frac{\delta}{ 2^{n+1} }.\]
\end{lemma}
\begin{proof}
By Theorem~\ref{thm:cut} there is a partition such that $A = A_0 \cup A_1$ such that $\opt(S) = |N(A_0) \cap S| + \sum_{a \in A_1} \sigma C_a$.  Now write $|N(A_0)\cap S| = \sum_{i \in S} Y_i$, where $Y_i = \mathbf{1}_{\{i \in N(A_0)\}}$.  Computing the expectation of this sum we have $\E[\sum_{i \in S} Y_i] = \sigma |N(A_0)|$.
Now we have
\[ \begin{split}
    &\Pr[\opt(S) < \sigma \opt - \eps^2 \sigma m] \\
    & \leq \Pr\left[ \sum_{i \in S} Y_i \leq \sigma |N(A_0)| - \eps^2 \sigma m \right]  \\
    & \leq \exp\left(-\eps^2 \sigma \frac{m}{2} \right) \leq \frac{\delta}{2^{n+1}}
\end{split}
\]
where we use the condition on $m$ in the last step.
\end{proof}

Now we can prove the main result about Algorithm~\ref{alg:weights_random_order}.

\begin{proof}[of Theorem~\ref{thm:main_thm}]
Our goal is to show that the properties in Lemma~\ref{lem:main_lemma} hold, then apply its conclusion.  Recall that we assume $m =\Omega(\frac{n^2}{\sigma \epsilon^2}\log(\frac{n}{\delta}))$ and that $\opt \geq \epsilon m$.  For the first property, from Lemma~\ref{lem:property1}, we have that for each $\alpha \in \cA(T,\epsilon)$ and $a \in A$
\[
\Pr\left[R_a(\alpha,S) >  \sigma R_a(\alpha) + \epsilon^2\sigma  \frac{ m}{n} \right] \leq \frac{\delta}{2n|\cA(T,\epsilon)|}.
\]
Note that $|\cA(T,\eps)| = T^n = (O\left(\frac{1}{\epsilon^2}\log(\frac{n}{\epsilon}) \right))^n$.  Union bounding over all $\alpha \in \cA(T,\eps)$ and $a\in A$, we get that this holds for the particular $\alpha$ computed by the algorithm and also we can sum the inequalities to get
\[
\Pr\left[\sum_a R_a(\alpha,S) > \sigma \sum_a R_a(\alpha) + \epsilon^2 \sigma m\right] \leq \frac{\delta}{2}
\]
showing the first property.  For the second property, we apply Lemma~\ref{lem:property2} and union bound over all partitions of $A$ into $A_0 \cup A_1$, of which there are at most $2^n$, yielding
\[
\Pr[\opt(S) < \sigma \opt - \epsilon^2 \sigma m] \leq \frac{\delta}{ 2}
\]
showing the second property.  Applying Lemma~\ref{lem:main_lemma} proves the theorem.
\end{proof}
\section{The Proof for Theorem~\ref{thm:robust}}\label{app:roubust}

\begin{proof}
	The basic framework of the proof is similar with the proof in~\cite{lavastida2020learnable}. Let $\val(\alpha,\cI)$ be the objective value obtained by using weights $\alpha$ proportionally on instance $\cI$. Thus, our main goal is to prove $\val(\predw,\cI) \geq (1-\epsilon)\opt - 2\eta$. 
	According to the property of the proportional weights~\cite{DBLP:journals/ior/AgrawalWY14}, if we add a dummy source $s$ adjacent to all impressions and dummy sink $t$ adjacent to all advertisers on the graph of instance $\predI$, there exists a vertex $s$-$t$ cut $\cC$ whose value $C(\cC,\predI)$ is at most $(1+\epsilon)\val(\predw,\predI)$. 
	As mentioned in Appendix~\ref{sec:proofs}, the cut $\cC$ is formed by $N(A_0)\cup A_1$, where $A_0$ and $A_1$ is two partitions of advertiser set $A$.
	Use $C(\cC,\cI)$ to represent the value of the cut $\cC$ in instance $\cI$.
	
	We first build the relationship between the performances of weights $\predw$ in the two instances:
	\begin{equation}\label{eq:robust_eq_1}
	\val(\predw,\cI) \geq \val(\predw,\predI) - \eta,
	\end{equation}
	and then prove the values of cut $\cC$ in $\predI$ and $\cI$ are also close:
	\begin{equation}\label{eq:robust_eq_2}
	C(\cC,\predI)\geq C(\cC,\cI) - \eta.
	\end{equation}
	Since $ C(\C,G)$ is the upper bound of $\opt$, the theorem can be proved directly by Eq.~\eqref{eq:robust_eq_1} and Eq.~\eqref{eq:robust_eq_2}.
	
    Note that we can assume that the graph connections in $\predI$ and $\cI$ are the same and the difference is the capacity of each vertex.
    Create a new instance $\cI'$ based on $\predI$ and $\cI$, where the connection is the same and the capacity of each vertex is the minimum value of its capacity in these two instance. Let $C_v(\cI)$ be the capacity of vertex $v$ in instance $\cI$. Namely, for each vertex $v\in \cI'$, we have
    $C_v(\cI') = \min \{C_v(\cI),C_v(\predI)\}$. Thus, we have
    \[ \sum_i|C_i(\cI') -C_i(\predI)| + \sum_a|C_a(\cI') -C_a(\predI)| \leq \eta, \]
    indicating that with the same weights, if the capacity of each vertex $v$ increases from $C_v(\cI')$ to $c_v(\predI)$, the objective value increases at most $\eta$. In other words, 
    \[ \val(\predw,\cI') \geq \val(\predw,\predI) -\eta. \] 
	Since the capacity of each vertex in $\cI$ is no less than that in $\cI'$, we know
	\[ \val(\predw,\cI) \geq \val(\predw,\cI') \geq \val(\predw,\predI) -\eta, \]
	completing the proof of Eq.~\eqref{eq:robust_eq_1}.

	Eq~\eqref{eq:robust_eq_2} can also be proved in the same way by analyzing the capacity of the cut in $\cI'$ and comparing this to the capacities in $\predI$ and $\cI$, respectively. Doing so yields the following chain of inequalities:
	\[ C(\C,\predI) \geq C(\C,\cI') \geq C(\C,\cI) - \eta.\]
	This now completes the proof as argued above.
\end{proof} 

\section{Additional Experimental Results}\label{sec:exp_app}
\begin{figure*}[t]
    \centering
    \begin{subfigure}[t]{0.32\textwidth}
         \centering
         \captionsetup{justification=centering}
	     \includegraphics[width=\linewidth]{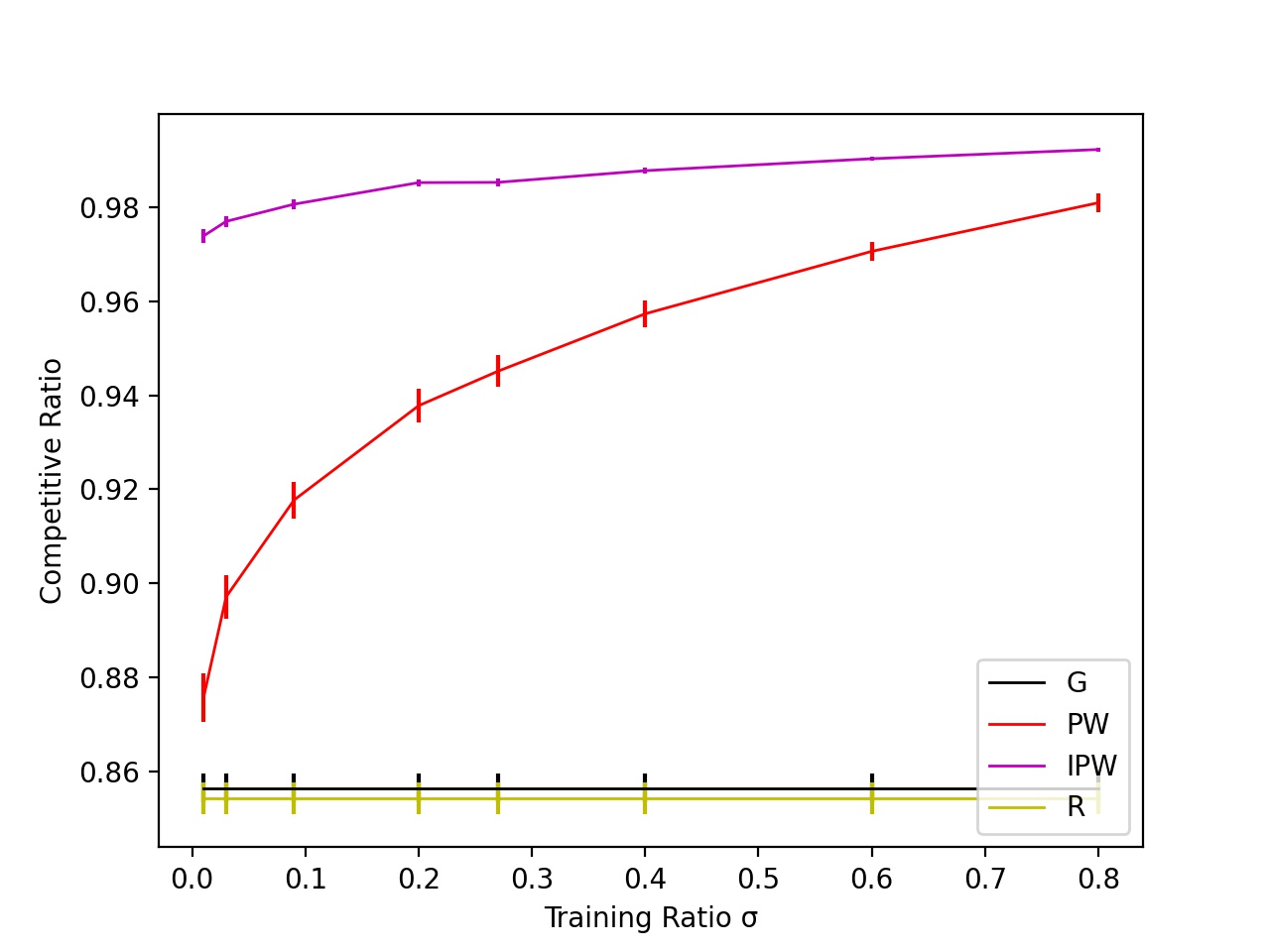}
         \caption{Random Quota}
         \label{fig:Random_Capacity_mode1_One_Day}
     \end{subfigure}
     \begin{subfigure}[t]{0.32\textwidth}
         \centering
         \captionsetup{justification=centering}
	     \includegraphics[width=\linewidth]{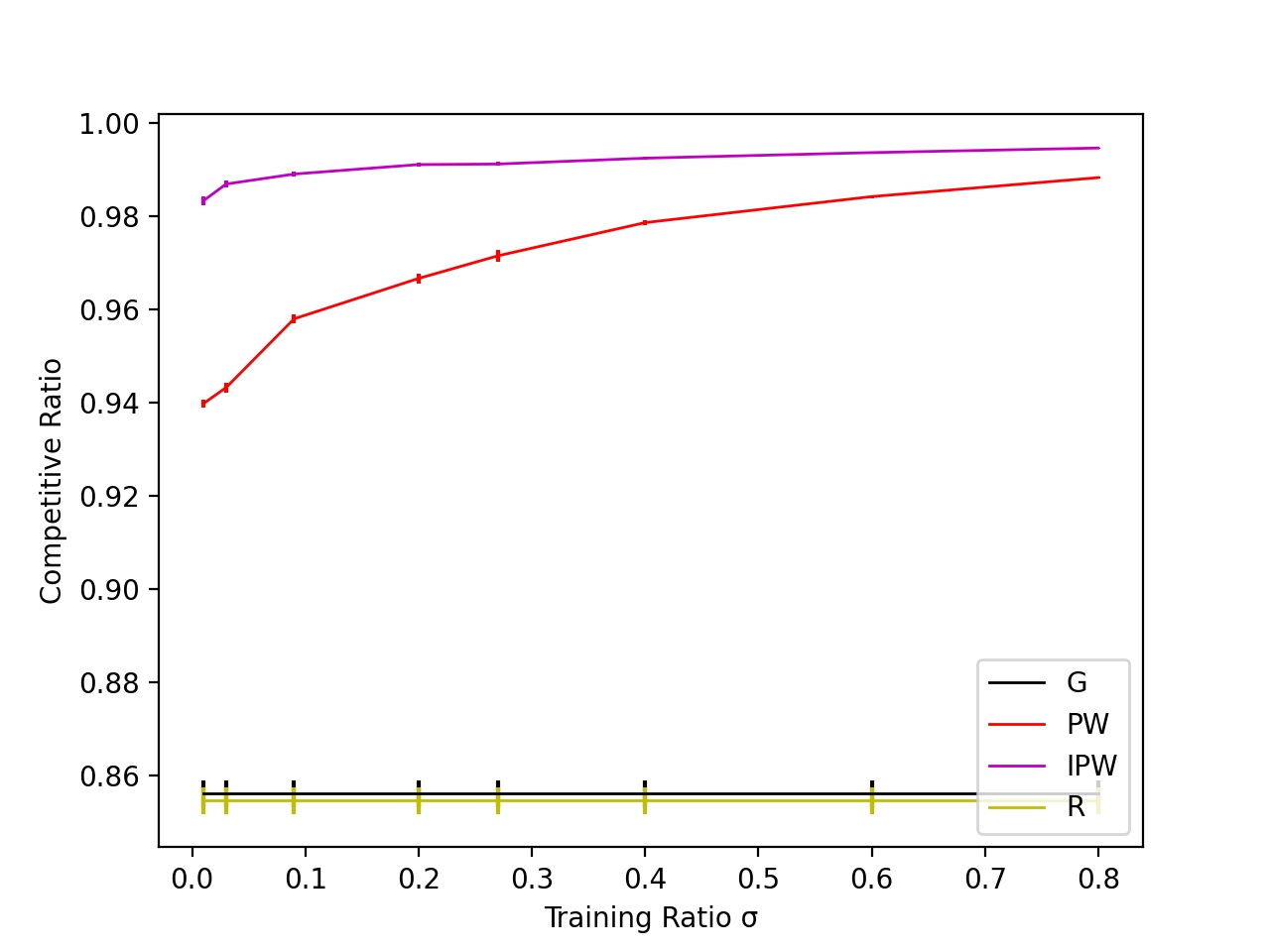}
         \caption{Max-min Quota}
         \label{fig:Min_Max_mode1_One_Day}
     \end{subfigure}
     \begin{subfigure}[t]{0.32\textwidth}
         \centering
         \captionsetup{justification=centering}
	     \includegraphics[width=\linewidth]{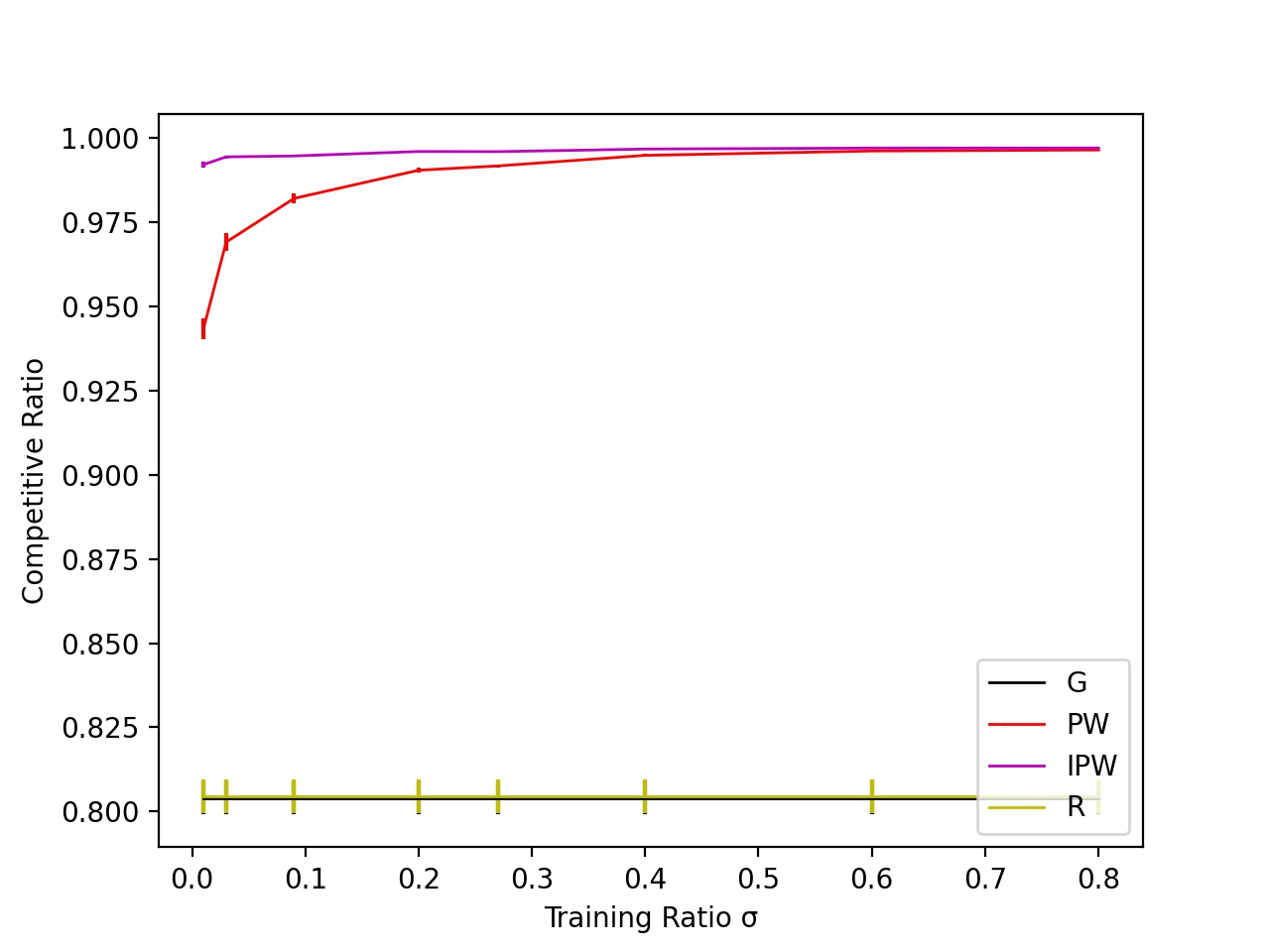}
         \caption{Least-degree Quota}
         \label{fig:Least_degree_mode1_One_Day}
     \end{subfigure}
    \caption{Competitive Ratios of different algorithms in the $C_i$-descending order, plotted as a function of the training ratio.}
    \label{fig:Performance_mode1_One_Day}
\end{figure*}

\begin{figure*}[t]
    \centering
    \begin{subfigure}[t]{0.32\textwidth}
         \centering
         \captionsetup{justification=centering}
	     \includegraphics[width=\linewidth]{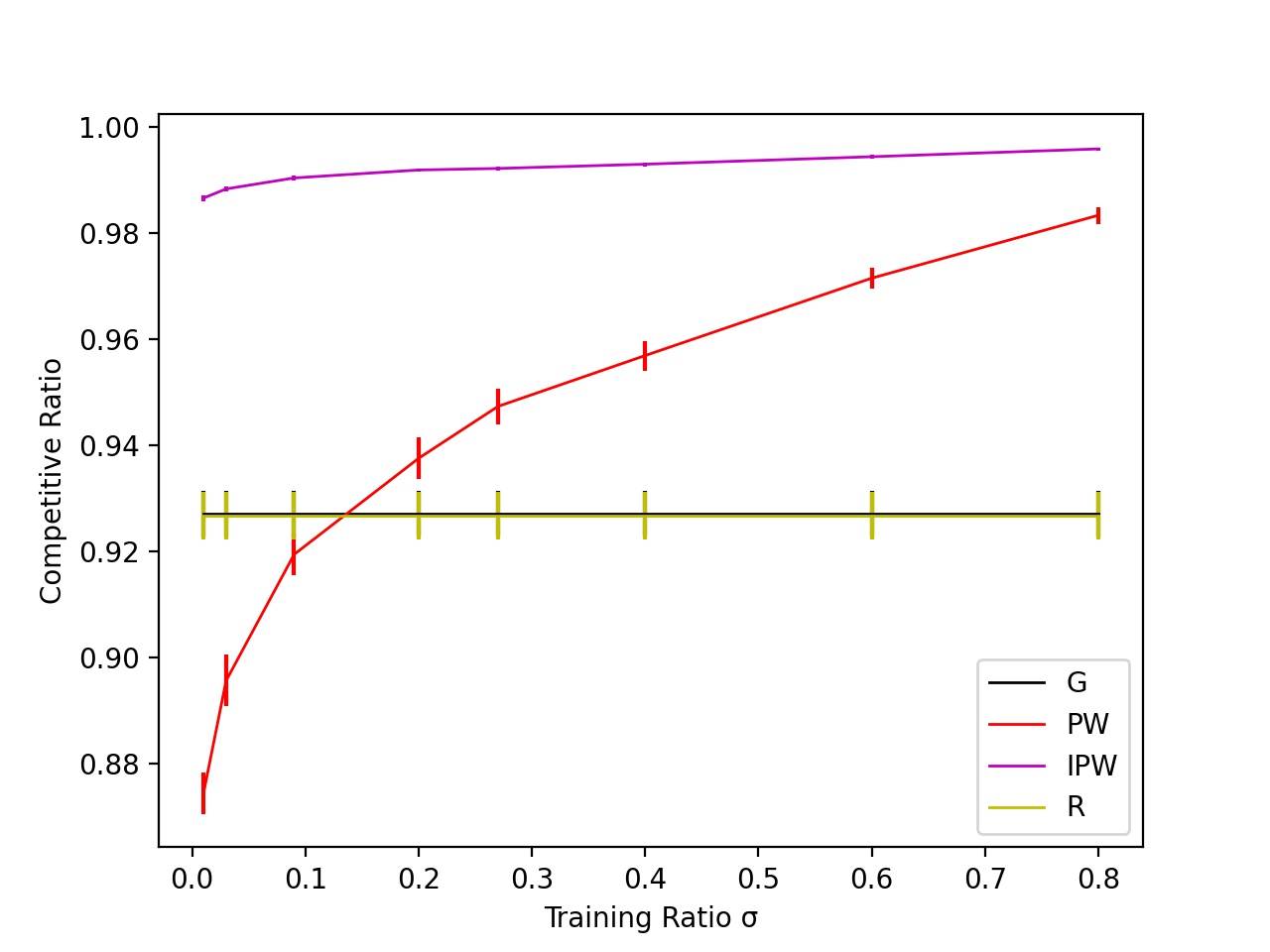}
         \caption{Random Quota}
         \label{fig:Random_Capacity_mode2_One_Day}
     \end{subfigure}
     \begin{subfigure}[t]{0.32\textwidth}
         \centering
         \captionsetup{justification=centering}
	     \includegraphics[width=\linewidth]{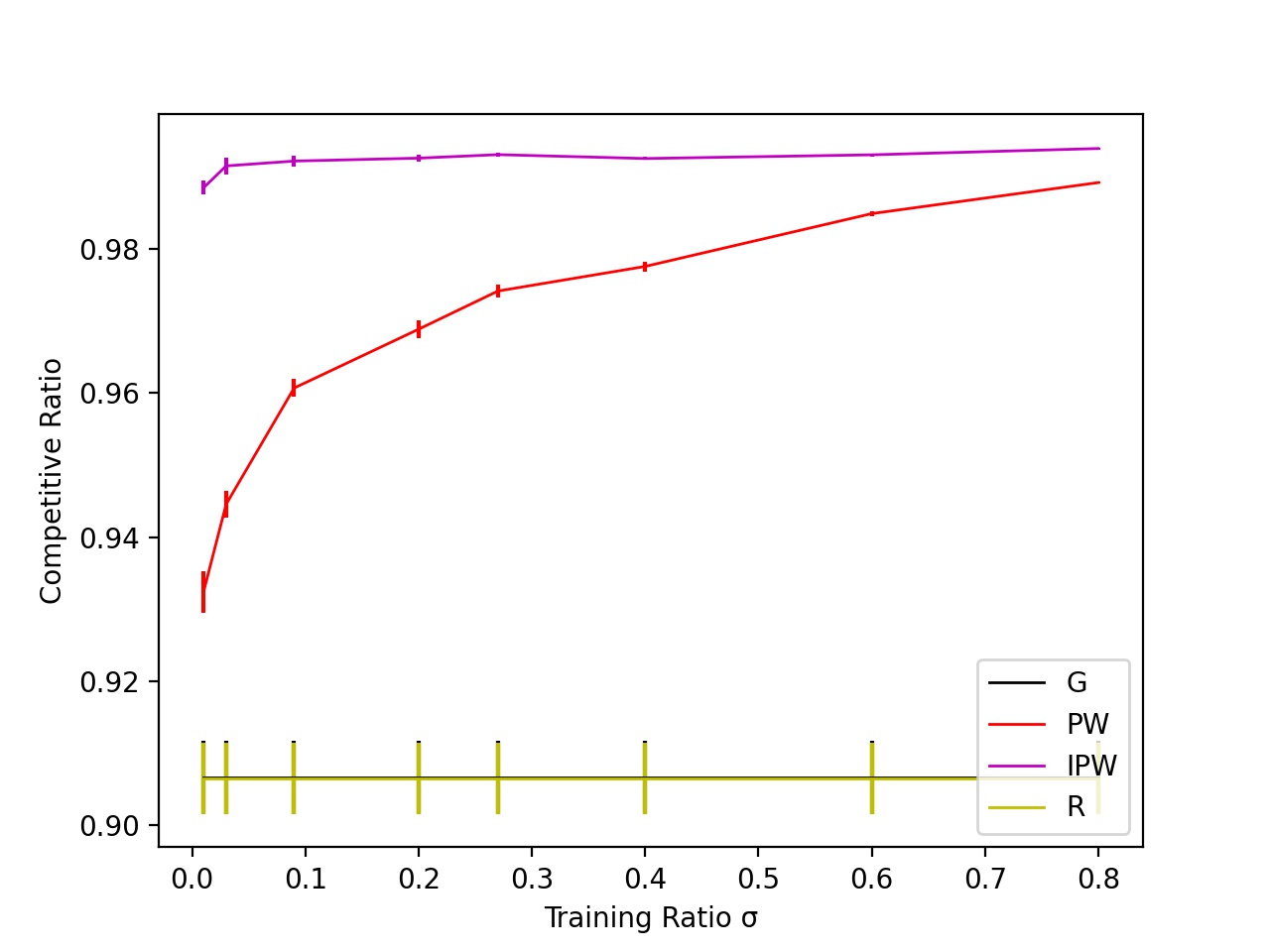}
         \caption{Max-min Quota}
         \label{fig:Min_Max_mode2_One_Day}
     \end{subfigure}
     \begin{subfigure}[t]{0.32\textwidth}
         \centering
         \captionsetup{justification=centering}
	     \includegraphics[width=\linewidth]{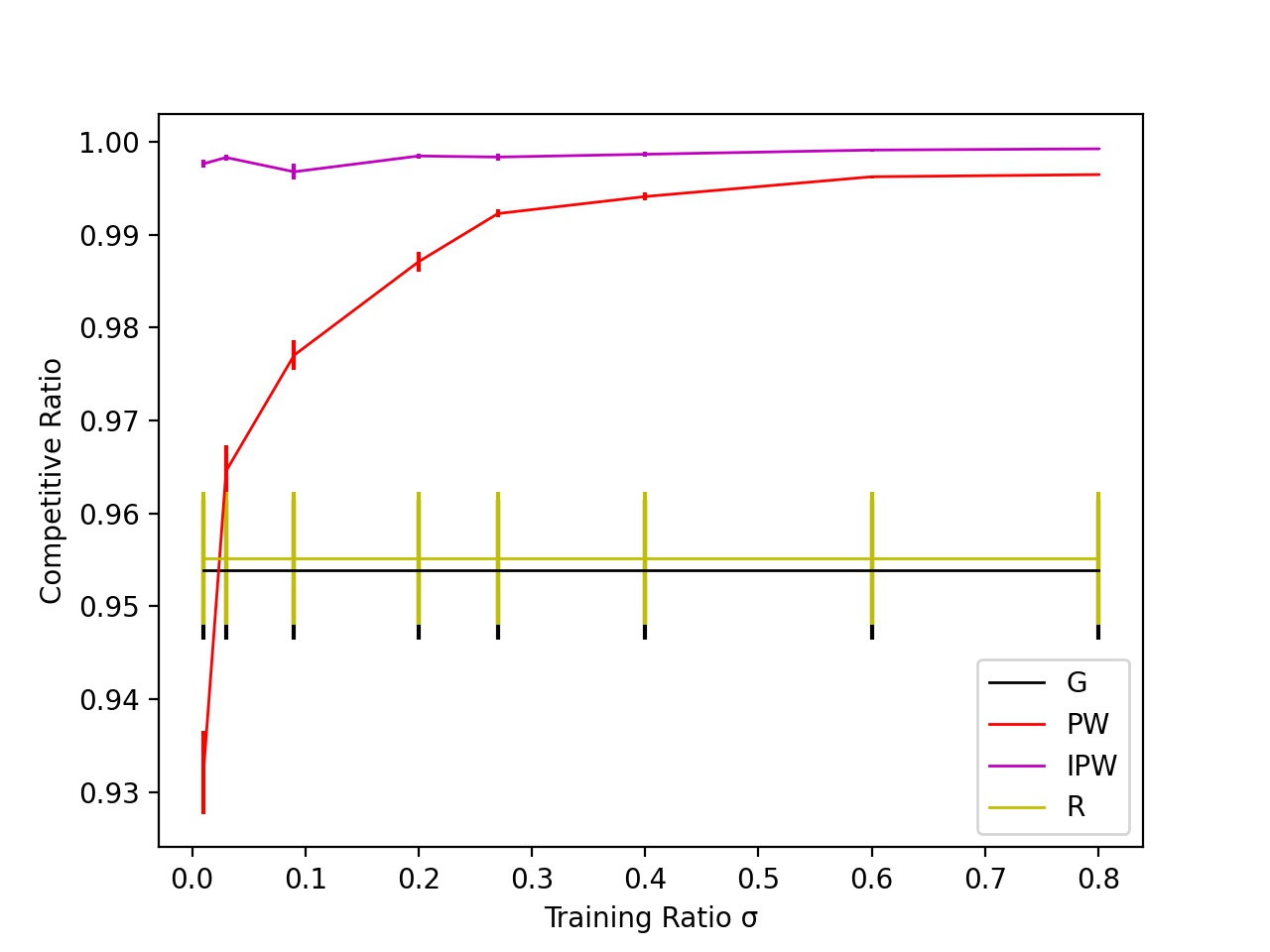}
         \caption{Least-degree Quota}
         \label{fig:Least_degree_mode2_One_Day}
     \end{subfigure}
    \caption{Competitive Ratios of different algorithms in the $C_i$-ascending order, plotted as a function of the training ratio.}
    \label{fig:Performance_mode2_One_Day}
\end{figure*}

\begin{figure*}[t]
    \centering
    \begin{subfigure}[t]{0.32\textwidth}
         \centering
         \captionsetup{justification=centering}
	     \includegraphics[width=\linewidth]{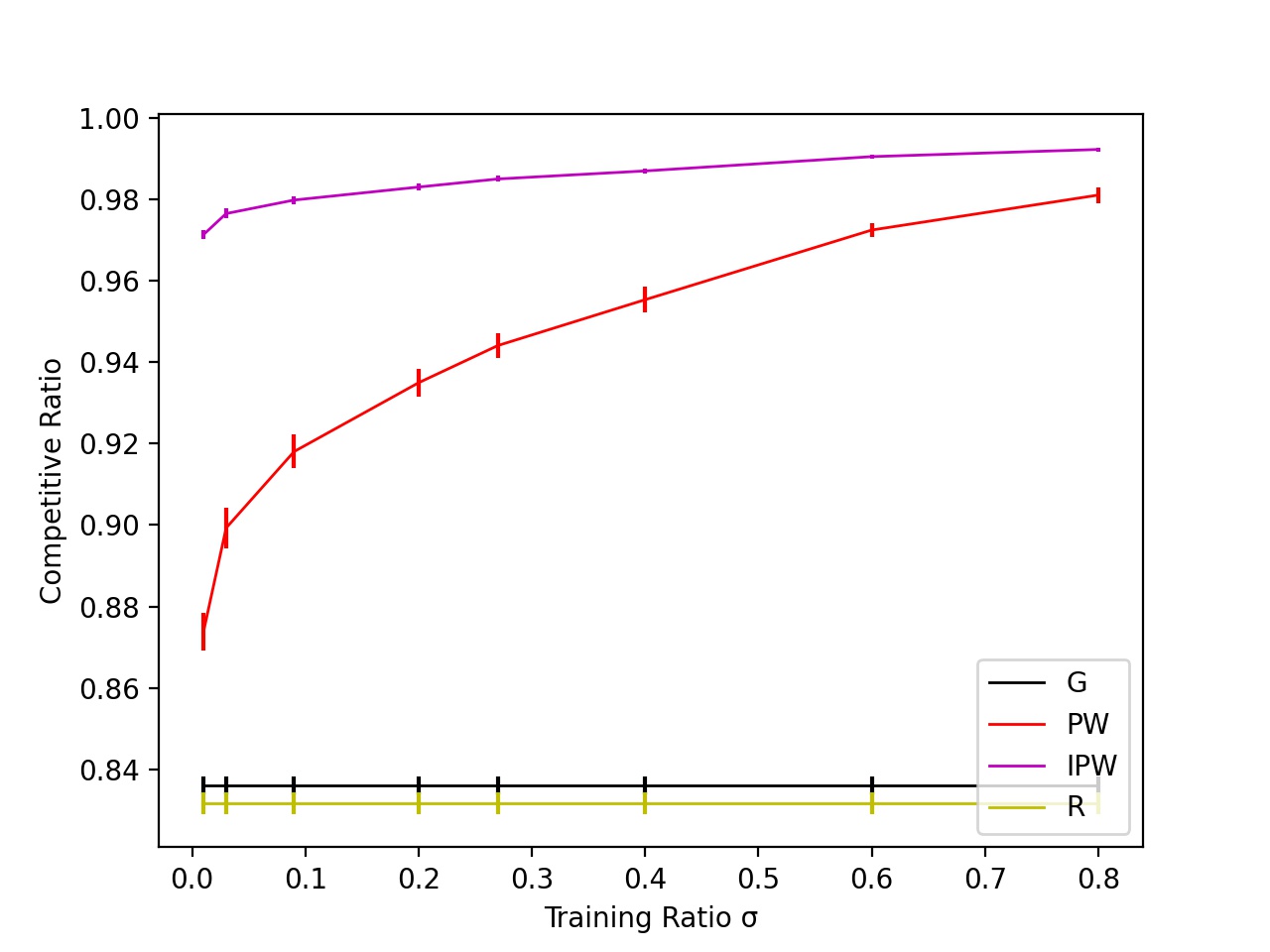}
         \caption{Random Quota}
         \label{fig:Random_Capacity_mode3_One_Day}
     \end{subfigure}
     \begin{subfigure}[t]{0.32\textwidth}
         \centering
         \captionsetup{justification=centering}
	     \includegraphics[width=\linewidth]{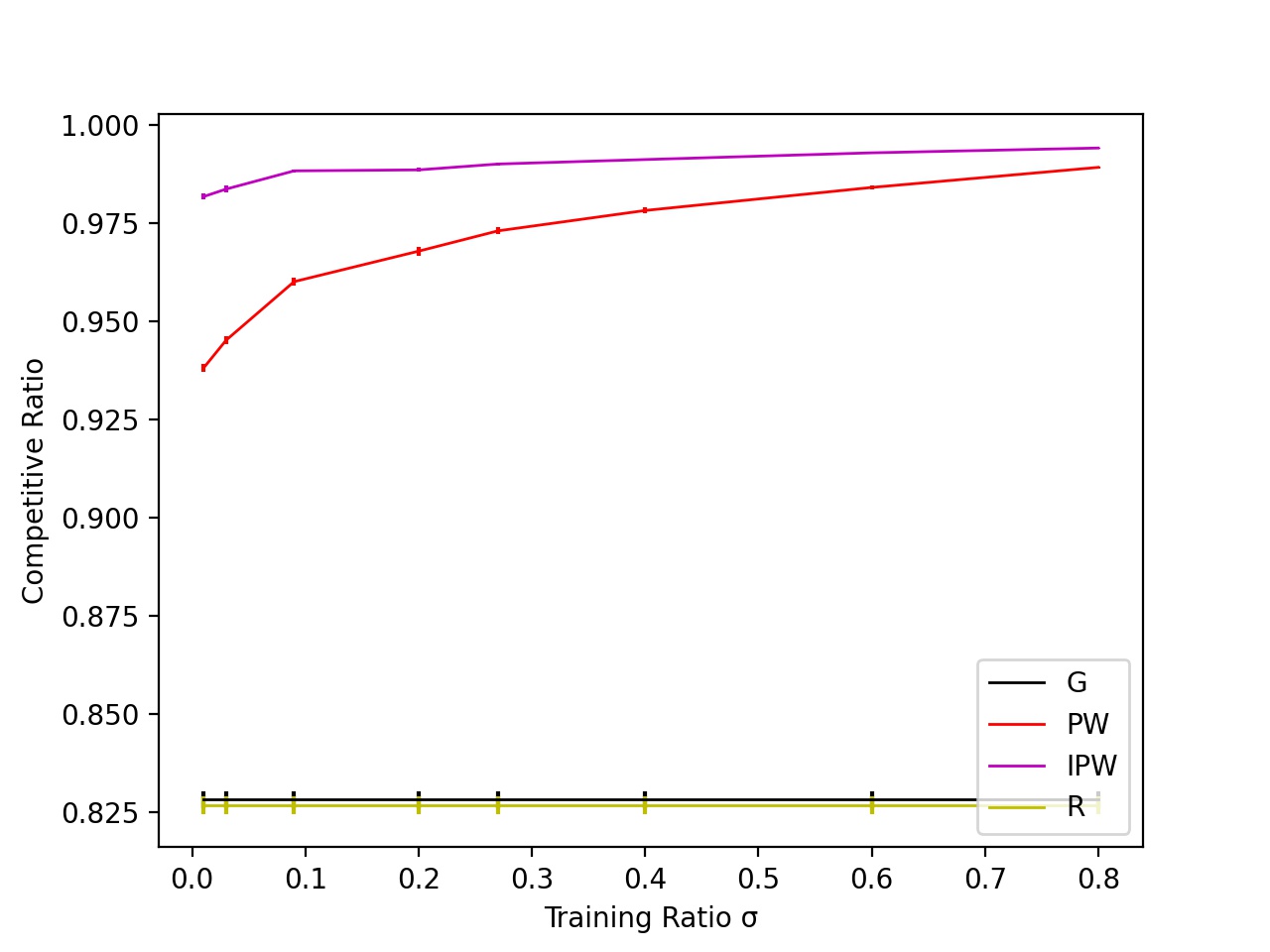}
         \caption{Max-min Quota}
         \label{fig:Min_Max_mode3_One_Day}
     \end{subfigure}
     \begin{subfigure}[t]{0.32\textwidth}
         \centering
         \captionsetup{justification=centering}
	     \includegraphics[width=\linewidth]{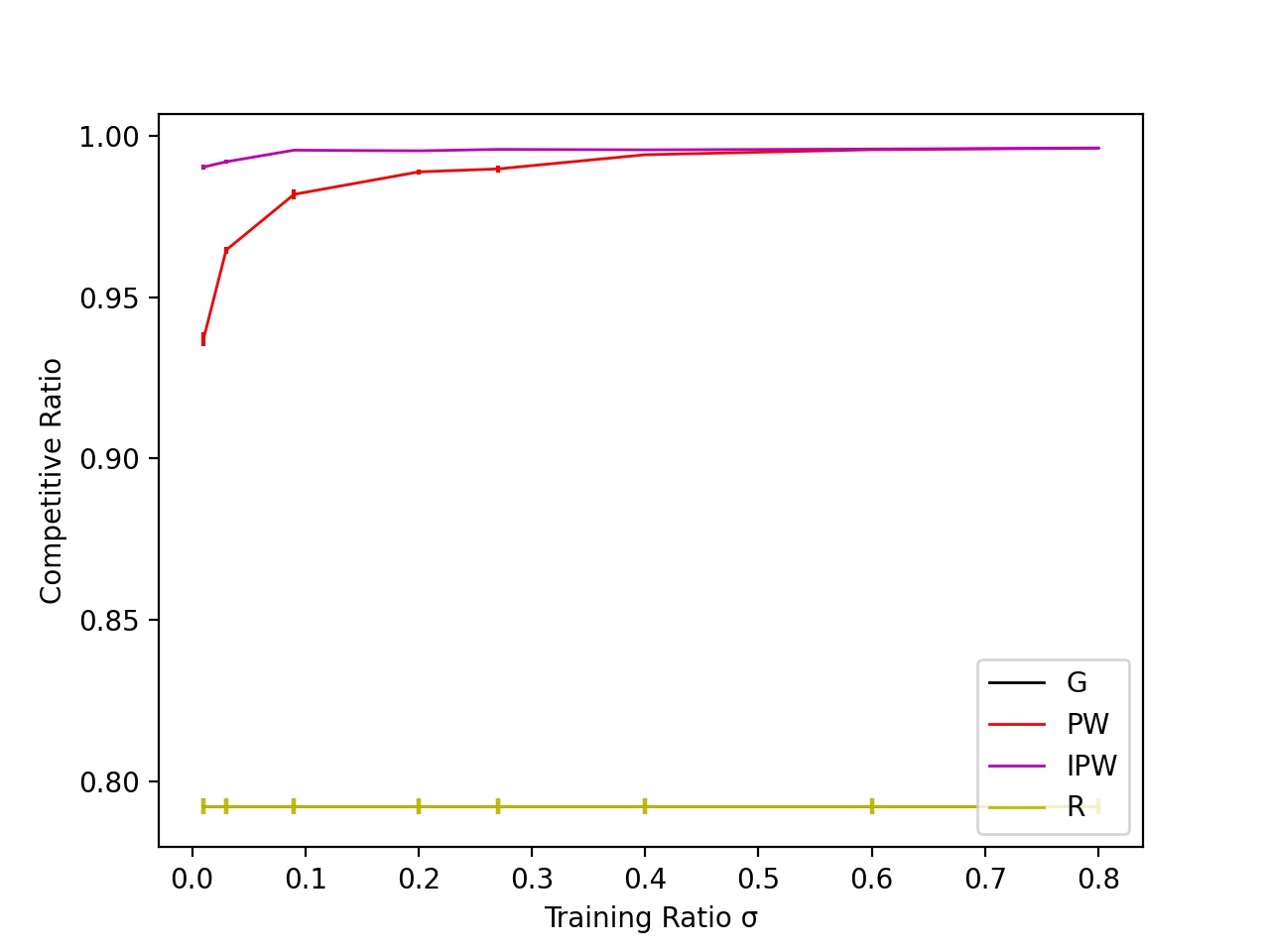}
         \caption{Least-degree Quota}
         \label{fig:Least_degree_mode3_One_Day}
     \end{subfigure}
    \caption{Competitive Ratios of different algorithms in the $C_a$-descending order, plotted as a function of the training ratio.}
    \label{fig:Performance_mode3_One_Day}
\end{figure*}

\begin{figure*}[t]
    \centering
    \begin{subfigure}[t]{0.32\textwidth}
         \centering
         \captionsetup{justification=centering}
	     \includegraphics[width=\linewidth]{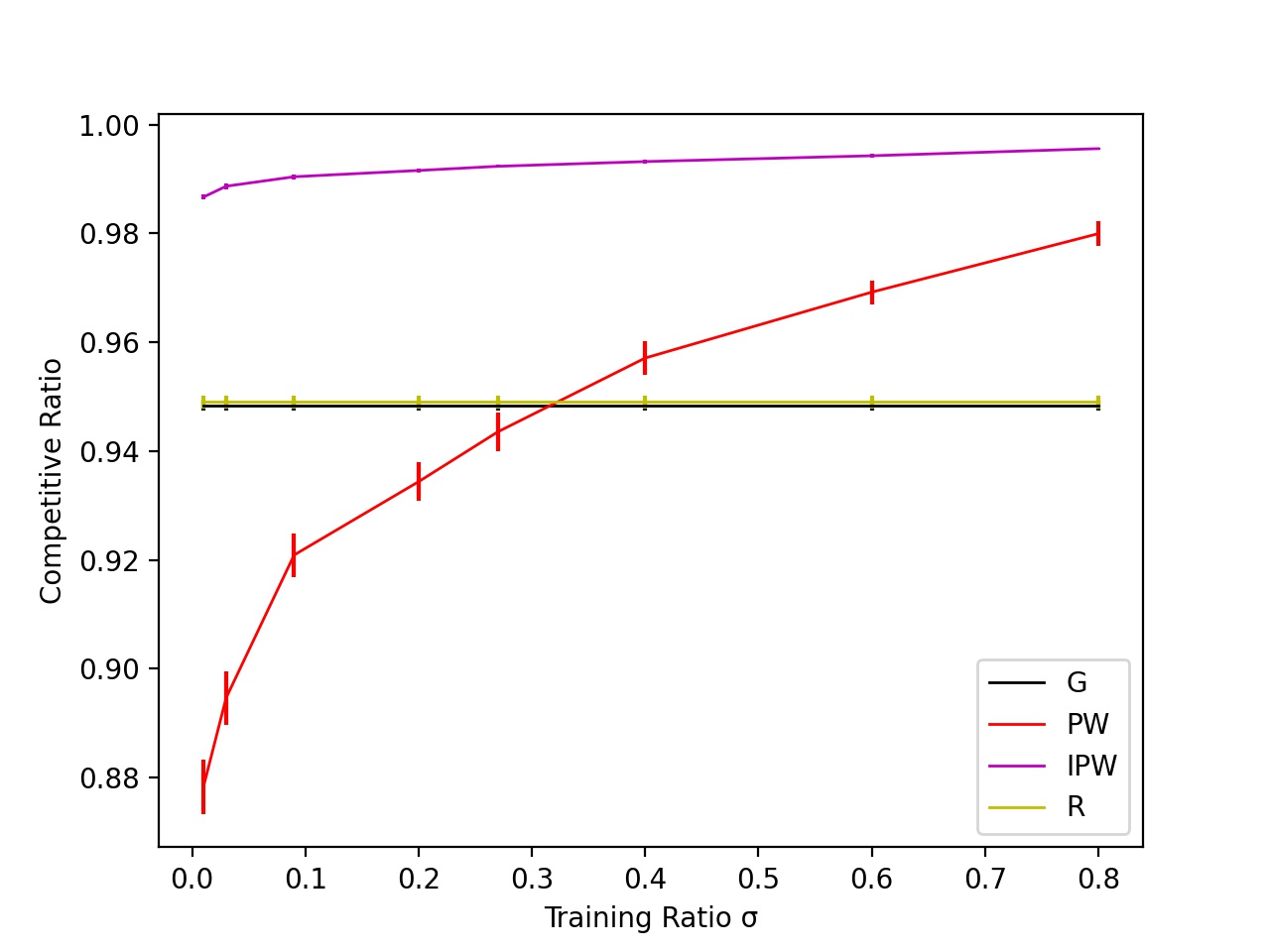}
         \caption{Random Quota}
         \label{fig:Random_Capacity_mode4_One_Day}
     \end{subfigure}
     \begin{subfigure}[t]{0.32\textwidth}
         \centering
         \captionsetup{justification=centering}
	     \includegraphics[width=\linewidth]{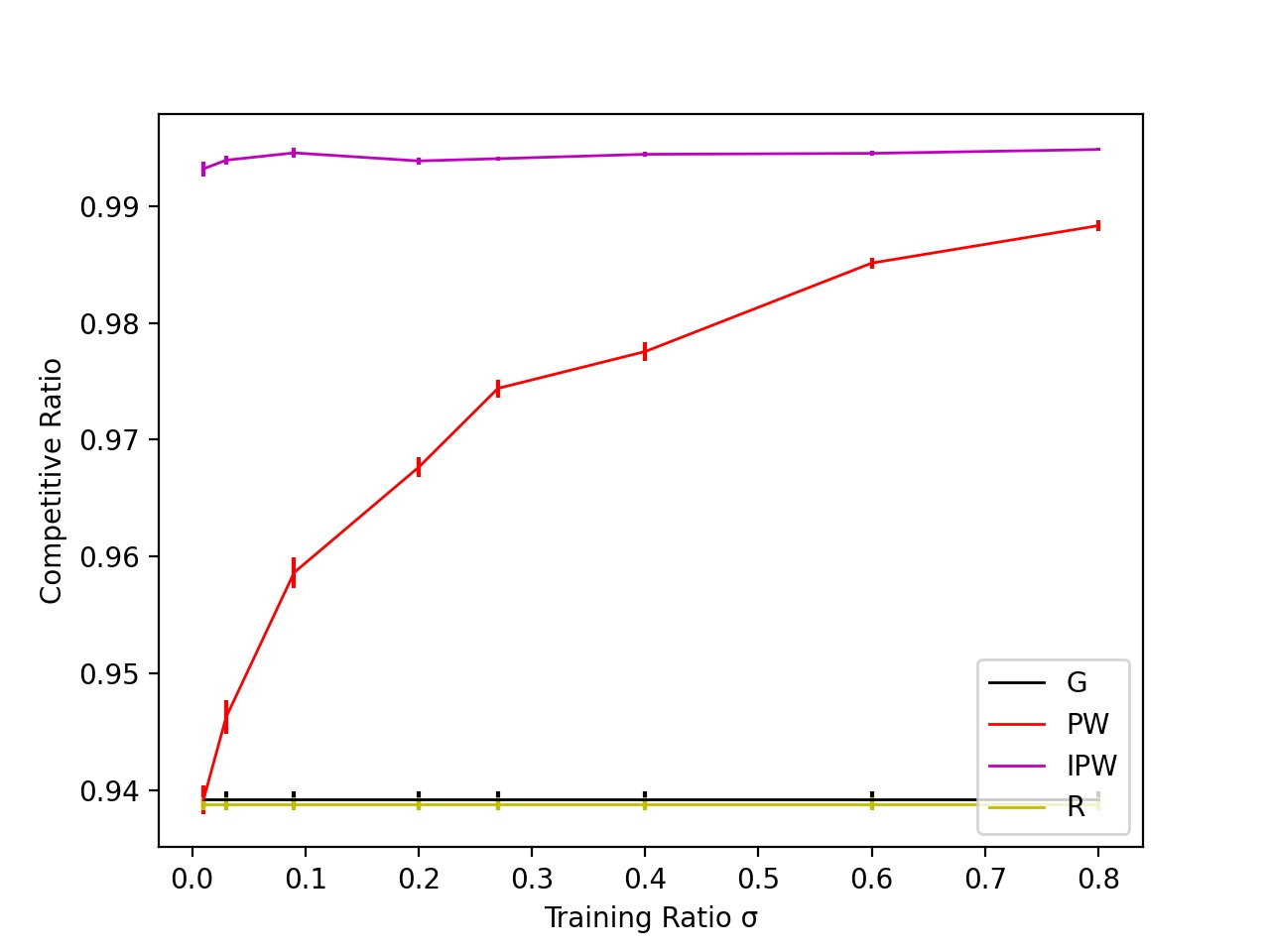}
         \caption{Max-min Quota}
         \label{fig:Min_Max_mode4_One_Day}
     \end{subfigure}
     \begin{subfigure}[t]{0.32\textwidth}
         \centering
         \captionsetup{justification=centering}
	     \includegraphics[width=\linewidth]{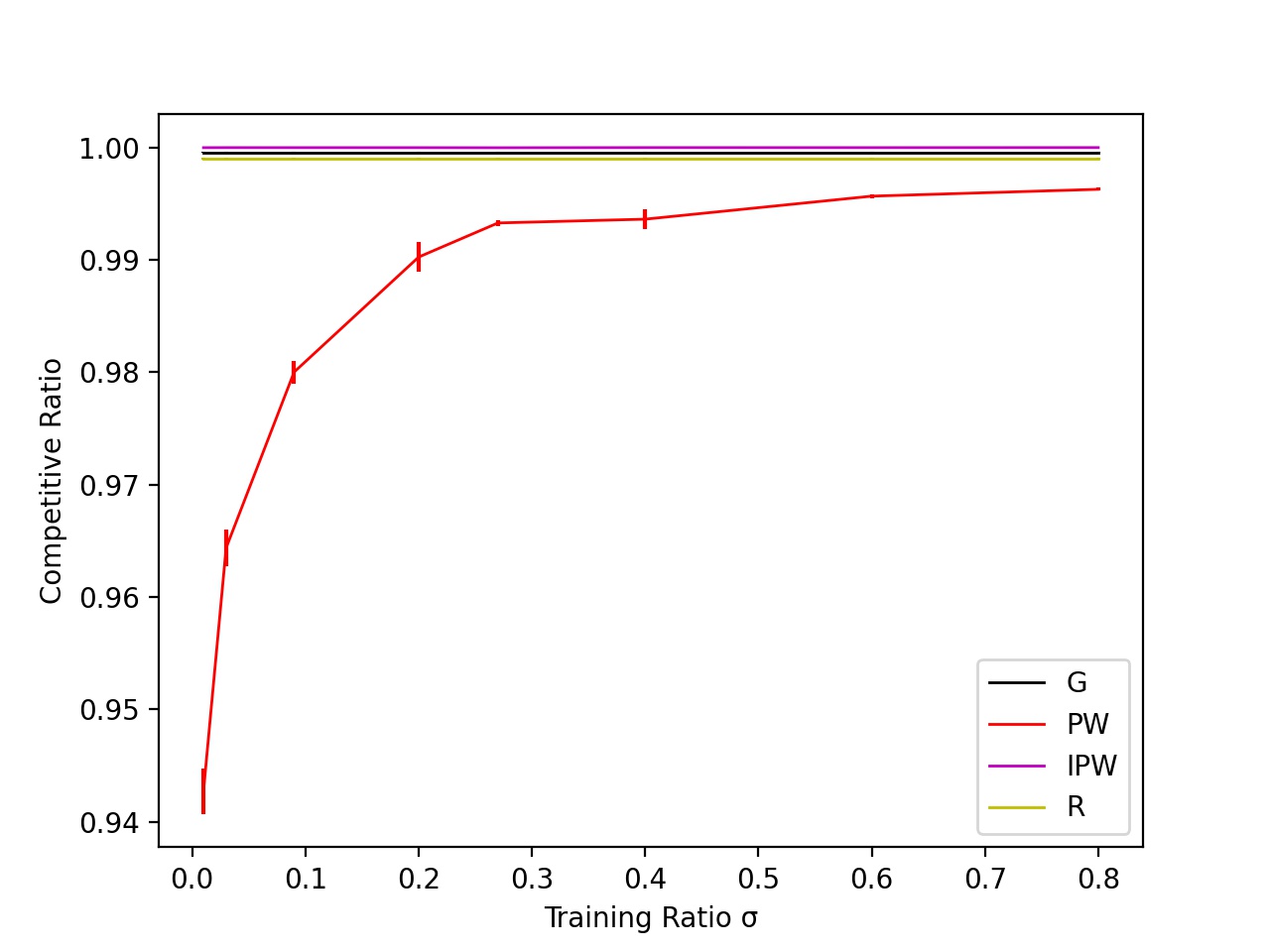}
         \caption{Least-degree Quota}
         \label{fig:Least_degree_mode4_One_Day}
     \end{subfigure}
    \caption{Competitive Ratios of different algorithms in the $C_a$-ascending order, plotted as a function of the training ratio.}
    \label{fig:Performance_mode4_One_Day}
\end{figure*}

\begin{figure*}[t]
    \centering
    \begin{subfigure}[t]{0.4\textwidth}
         \centering
         \captionsetup{justification=centering}
	     \includegraphics[width=\linewidth]{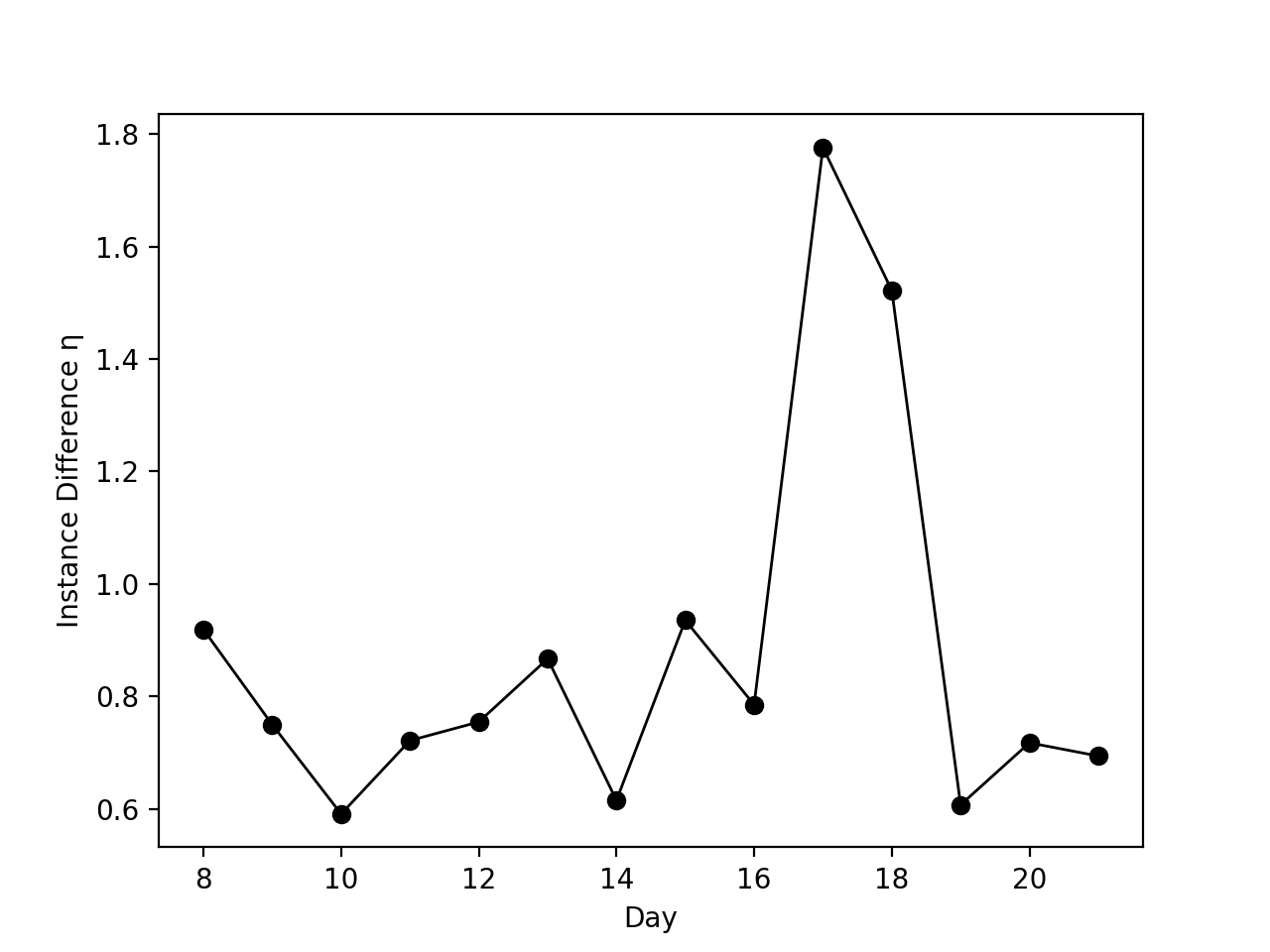}
         \caption{Max-min Quota}
         \label{fig:max-min_l1_norm}
     \end{subfigure}
     \begin{subfigure}[t]{0.4\textwidth}
         \centering
         \captionsetup{justification=centering}
	     \includegraphics[width=\linewidth]{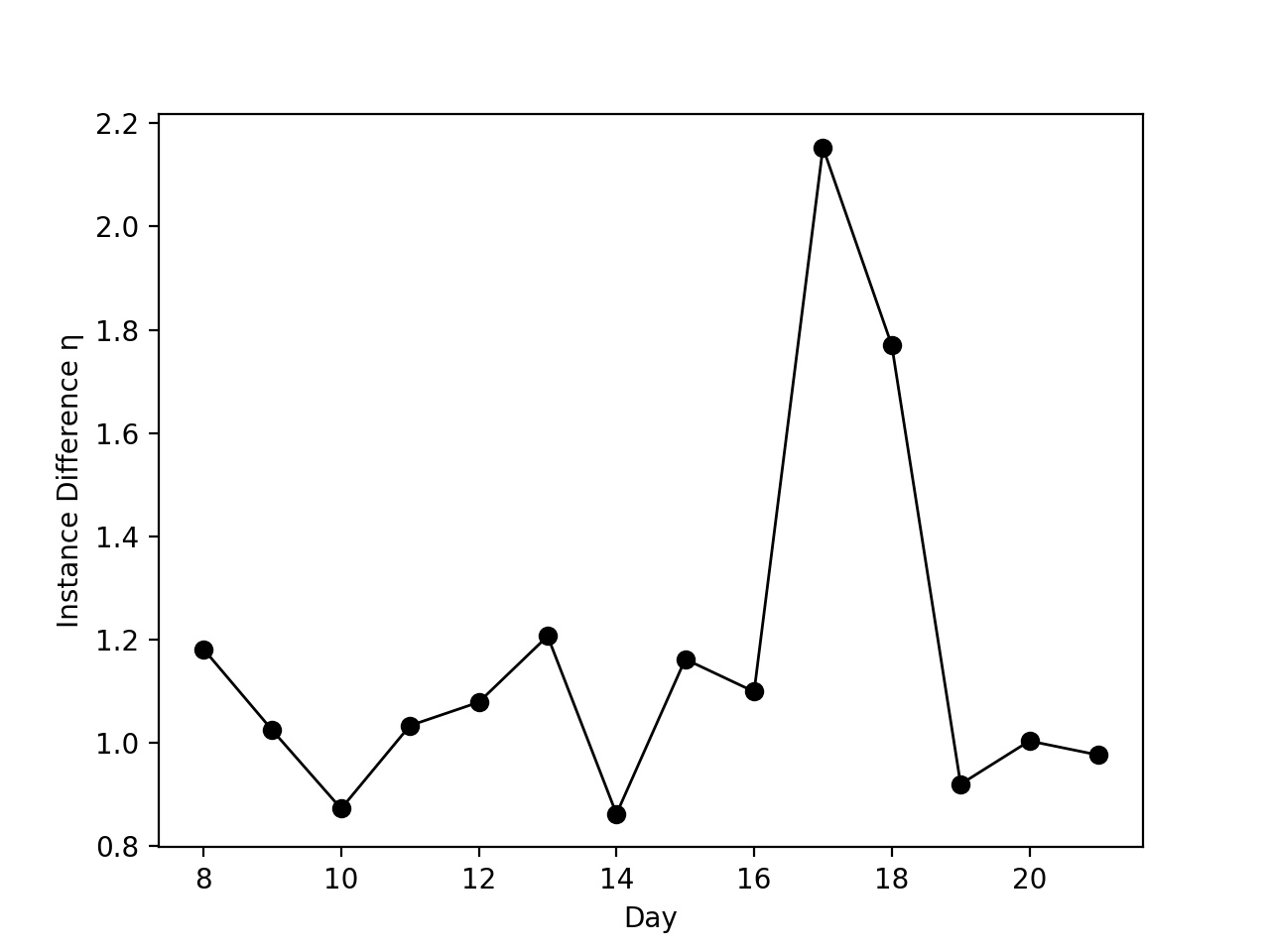}
         \caption{Least-degree Quota}
         \label{fig:least-degree_l1_norm}
     \end{subfigure}
    \caption{The instance difference $\eta$ between day $i$ and day $i-1$ for $i\in[8,21]$ under two capacity settings, where the instance difference is the $\ell_1$ norm between the impression vectors (normalized) of these two instances plus the $\ell_1$ norm between two advertiser vectors (normalized). Note the different scales in the Y-axis.}
    \label{fig:instance_difference}
\end{figure*}

This section shows more results in the experiments. Recall that we used five arrival orders in the learnability experiments and show the performance in the random order and the worst performance of each algorithm among these five orders. Now we present the performance of different algorithms in the remaining four impression arrival orders (other than random) in Fig.~\ref{fig:Performance_mode1_One_Day}~-~\ref{fig:Performance_mode4_One_Day}. For the robustness experiments in Fig.~\ref{fig:Performance_Different_Day}, we give the instance differences under different capacity settings in Fig.~\ref{fig:instance_difference}.
In most places, when the instance difference $\eta$ decreases, algorithm PW$\_1$ tends to increase. 

\end{document}